\documentclass[10pt,final,doubleside]{IEEEtran}
\normalsize
\IEEEoverridecommandlockouts

\usepackage{booktabs}
\usepackage[table,xcdraw]{xcolor}
\usepackage{url}
\usepackage{cite}
\usepackage{amsmath,amssymb,amsfonts}
\usepackage{algorithmic}
\usepackage{graphicx}
\usepackage{textcomp}
\usepackage{xcolor}
\usepackage{pifont,bm,multicol,amsfonts,amsmath,color,amssymb,graphicx, epsfig,cite,psfrag,subfigure,algorithm,enumerate,stfloats,algorithm,algorithmic,epstopdf}
\usepackage{caption}
\usepackage{float}
\usepackage[normalem]{ulem}
\usepackage{comment}
\usepackage{soul}

\usepackage{balance}
\usepackage{cancel}
\usepackage{siunitx}
 \newtheorem{observation}{\underline{Observation}}

\newtheorem{theorem}{\noindent\underline{Theorem}}
\newtheorem{lemma}{\noindent\underline{Lemma}}

\newtheorem{corollary}{\underline{Corollary}}

\newtheorem{proposition}{\underline{Proposition}}

\def\BibTeX{{\rm B\kern-.05em{\sc i\kern-.025em b}\kern-.08em
    T\kern-.1667em\lower.7ex\hbox{E}\kern-.125emX}}

\begin{document}

\bibliographystyle{ieeetr}

\title{
The Throughput Gain of Hypercycle-level Resource Reservation for Time-Triggered Ethernet
}

\author{
Peng Wang, Suman Sourav, Binbin Chen, \emph{Member, IEEE}, Hongyan Li, \emph{Member, IEEE}, Feng Wang, Fan Zhang\\ 


\thanks{P. Wang, S. Sourav, B. Chen and F. Wang are with the pillar of Information Systems Technology and Design, Singapore University of Technology and Design, 487372, Singapore (Email: \{peng\_wang2, binbin\_chen, suman\_sourav, feng2\_wang\}@sutd.edu.sg),(Corresponding author: P. Wang).}

\thanks{ H. Li is with the State Key Laboratory of Integrated Services Networks, Xidian University, Xi'an 710071, China (Email: hyli@xidian.edu.cn).}

\thanks{F. Zhang is with the College of Computer Science and Technology, Zhejiang University, Hangzhou 310058, China (Email:fanzhang@zju.edu.cn)}
}

\maketitle

\vspace{-10mm}

\begin{abstract}
Time-Triggered Communication is a key technology for many safety-critical systems, with applications spanning the areas of aerospace, industrial control, and automotive electronics. Usually, such communication is reliant on time-triggered flows, 
\textcolor{black}{with each flow consisting of periodic packets that need to be delivered from a source to a destination node. Each packet needs to reach its destination before its deadline. }
Different flows can have different cycle lengths.  
To achieve assured transmission of time-triggered flows, existing efforts 
 constrain the packets of a flow to be cyclically transmitted along the same path based on the flow's given cycle length. 
Under such Fixed Cyclic Scheduling (FCS),  reservation for flows with different cycle lengths can 
become incompatible over a shared link, 
limiting the total number of admissible flows. 
Considering the cycle lengths of different flows, a hyper-cycle has length equal to their least common multiple (LCM). It determines the time duration over which the scheduling compatibility of the different flows can be checked.
In this work, we propose a more flexible scheduling scheme called the Hypercycle-level Flexible Scheduling (HFS) scheme, where a flow's resource reservation can change over its different cycles within the same hypercycle.
HFS can significantly increase the number of admitted flows by providing more scheduling options while remaining perfectly compatible with existing Time-Triggered Ethernet system. We show that, (theoretically) the possible capacity gain provided by HFS over FCS can be unbounded. 
We formulate the time-triggered joint pathfinding and scheduling problem under HFS as an integer linear programming problem which we prove to be NP-Hard. To solve HFS efficiently, we further propose a least-load-first heuristic~(HFS-LLF), solving HFS as a sequence of shortest path problems. 
Extensive study under real-world settings shows that HFS admits up to $6\times$ the number of flows scheduled by FCS. Moreover, our proposed HFS-LLF can run $10^4\times$ faster than solving HFS using a generic solver. 

\end{abstract}

\IEEEpeerreviewmaketitle{}

\section{Introduction}
Time-triggered Ethernet (TTEthernet) is widely used to provide deterministic and periodic delivery of packets among sensors, controllers, and actuators in real-time systems, which are used in the areas of automotive electronics, aerospace, and industrial control.
As a deterministic, synchronized, and congestion-free network, all nodes in TTEthernet are synchronized by a global clock. 
 TTEthernet uses individual time slots over each link as a fine-grained resource scheduling unit, where a resource unit is allocated to a specific flow 
to avoid potential contention and hence to provide high-reliability and low-latency assurance~\cite{schweissguth2017ilp}. 
Here, a flow refers to a sequence of periodic packets between a given source and destination pair. In other words, each flow has one packet every cycle, and different flows can have different cycle lengths.  
The fine-grained resource unit allocation in TTEtherne eliminates collision and
reduces the queuing delay. In contrast, time-sensitive networks as in~\cite{Marek2020TCOM} allow a resource unit to be used by multiple pairs of source and destinations, leading to non-negligible queuing delay. 

\begin{figure*}[]
\centering
\includegraphics[scale=0.9]{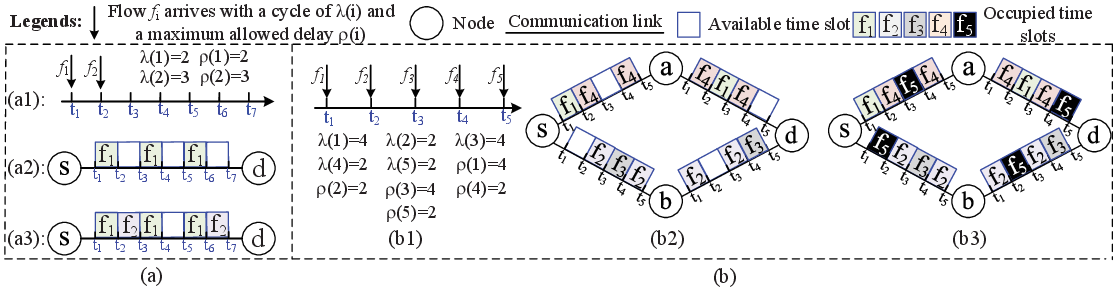}
\caption{ Examples of Fixed Cyclic Scheduling (FCS) and Hypercycle-level Flexible Scheduling (HFS).}
\label{fig:effects}
\end{figure*}

Several works (e.g., \cite{huang2021online,quan2020line,wang2023dmpf,Zhong2021,schweissguth2017ilp,yaoxu'24,Marek2020TCOM,yan2020injection,Zhang2022TII}) study the scheduling problem for TTEthernet. These works, however, all belong to the general category of {\it Fixed Cyclic Scheduling (FCS)}. 
Usually, each packet of a flow arrives in the TTEthernet cyclically and has a fixed maximum delay that determines the deadline by which the packet needs to be processed by the destination. 
 To ensure no jitter, i.e.,  the application layer at destination node of a flow can cyclically receive the packets, FCS schemes reserve a sequence  of periodic time slots over each link of a selected path.
Specifically, 
for each flow $f$ with a cycle length of $\lambda$ time slots, originating from source $s$ and destined for node $d$, FCS selects a sequence of links in TTEthernet to form a path from $s$ to $d$.  The packets of $f$ are periodically transmitted over each link in the selected path. For instance, for the network in  Fig. \ref{fig:effects}, if FCS selects the path $\{(s,a),(a,d)\}$ for a flow $f$ with a cycle of 2 time slots, it may transmit the first packet of $f$ over $(s,a)$ in the first time slot and  over edge $(a,d)$ in the second time slot. Next, the other packets will be periodically transmitted over $(s,a)$ in the 3rd, 5th, 7th... time slots and over $(a,d)$ in the 4th, 6th, 8th... time slots.
 Particularly, works \cite{schweissguth2017ilp,tvt23xiaolongwang,yaoxu'24} co-optimize the path selection and scheduling to exploit more scheduling options as compared to the works\cite{huang2021online,Marek2020TCOM,yan2020injection,Zhang2022TII} which separately conducts path selections and scheduling. On the other hand, studies\cite{yaoxu'24,Marek2020TCOM,krolikowski2021computercommunications,tvt24shi} seek to enable temporary storage at intermediate nodes for more schedule options in contrast to the efforts that adopt no-wait transmission at intermediate nodes. Moreover, references \cite{quan2020line,yan2020injection,tvt23xiaolongwang} study the best transmission time of the flow by allowing the flow to be temporarily stored at the source nodes as compared to other works \cite{pahlevan2019heuristic-sigbed,huang2021online} which transmit the data at the earliest time. Nonetheless,  flexible resource reservation across a flow's different periods has remained unexplored. 

FCS reserves the resources for a given flow following its cycle length. Hence, the resource reservation cycles under FCS for different flows can be different. This may result in conflict among flows with different cycle lengths, and limit the total number of flows that can be admitted.
  %
%
We illustrate how the difference between flows' cycle length can cause inefficiency in terms of the number of admitted flows by FCS, using an example as shown in Fig. \ref{fig:effects}(a).  
 Specifically, Fig. \ref{fig:effects} (a1) shows the time slots of the hypercycle for two flows $f_1$ and $f_2$ generated at node $s$ to reach node $d$, with cycle lengths of 2 and 3 time slots respectively.  The maximum allowed delay for each packet of flows $f_1$ and $f_2$ (to reach the destination $d$) is given to be 2 and 3 time slots respectively.  
Here, hypercycle refers to  
the least common multiple (LCM) of the cycles of all the flows in TTEthernet~\cite{huang2021online,quan2020line,wang2023dmpf,Zhong2021,schweissguth2017ilp,yaoxu'24,Marek2020TCOM,yan2020injection}. 
That is, the hypercycle for flow $f_1$ and $f_2$ is 6 time slots.
The maximum allowed delay, as determined by the application requirements, is the maximum delay, which the packet can experience, from the moment the packet arrives at TTEthernet to the moment it is received by the application layer at the destination nodes. The first packet of flows $f_1$ and $f_2$ arrive at time points $t_1$ and $t_2$ respectively. 
 Fig. \ref{fig:effects}~(a2) shows FCS schedules $f_1$ (with cycle length 2) to transmit packets within time slots $\tau_1 =[t_1,t_2], \tau_3 =[t_3,t_4]$ and $\tau_5=[t_5,t_6]$, and is
 unsuccessful in scheduling $f_2$. This is because FCS scheme only allows $f_2$ to be transmitted every 3 time slots over each link and as we can see the unoccupied slots do not meet that condition. 
This example highlights the fact that the source of inefficiency in scheduling multiple flows with different cycle lengths is potentially from the strict cyclic transmission over each link. Such strict cyclic transmission of packets of FCS over each selected link can be unnecessary and over-constrained. As long as each packet can arrive at the destination within its required delay (on or before its deadline),  the destination node can buffer the individual received packets within a flow according to the largest delay  experienced by the flow's packets. As such, even when the transmission is in a non-cyclic manner, the application layer at the destination can still receive packets with no jitter.  

With the above understanding, a natural question arises:
\textit{How much improvement can we achieve in terms of the number of admitted
flows when we relax the periodic transmission constraint while still satisfying the cyclic reception requirements of packets at destination nodes and satisfying the delay constraints?} To answer this question, we propose a {\it Hypercycle-level Flexible Scheduling (HFS)} scheme. The fundamental distinction between HFS and FCS 
lies in the degree of correlation of the schedule of different packets within the same flow. Specifically, HFS independently determines, for each packet of a flow within a hypercycle, both the path it will take and the packet transmission time on each link of the path, under the constraints that all packets arrive at its destination node no later than its deadline. In contrast, FCS computes the path and transmission time only for the first packet of the flow, subsequently transmitting all other packets in a periodic manner. Vividly speaking, in TTEthernet with a hypercycle of length $\Gamma$, 
 each original flow $f_i$, with a cycle of $\lambda(i)$, sends $\Gamma / \lambda(i)$ packets within one hypercycle. HFS considers these $\Gamma / \lambda(i)$ packets as independent packets originating from $\Gamma / \lambda(i)$ different flows, with each flow having the same cycle length of $\Gamma$. 
 This gives HFS more scheduling flexibility than FCS and potentially increases the number of admissible flows. 
We use Fig. \ref{fig:effects} to illustrate two potential sources of the performance gain of HFS over FCS.

Fig. \ref{fig:effects}(a) shows that HFS can outperform FCS by flexibly reserving time slots for different packets within a flow.  Recall that, FCS can only admit flow $f_1$ as shown in Fig. \ref{fig:effects}(a2), whereas HFS can successfully schedule both flows $f_1$ and $f_2$ as shown in Fig. \ref{fig:effects}(a3). Specifically,  the two packets of $f_2$ arrive at $t_2$ and $t_5$ respectively. The deadlines for the two packets to reach $d$ are $t_5$ and $t_8$, 
 respectively. 
 Considering 
 this, HFS delivers the two packets of $f_2$    to node $d$ at $\tau_2 = [t_2, t_3]$ and $\tau_6 = [t_6, t_7]$, respectively. 
 Notably, HFS can still achieve the periodic reception of packets at the application layer of the destination node. 
 That is, HFS can let the destination node temporarily hold selected packets so that the application layer can still receive all packets in a periodic manner. 

Performance gain can also come from reserving resources flexibly in different paths across different periods of a flow.
Fig. \ref{fig:effects}(b) gives a scheduling example where HFS flexibly reserving resources within a hypercycle admits more flows than FCS via flexibly selecting paths.
Fig. \ref{fig:effects}(b1) shows the arrival times of 5 flows in the first hypercycle. The source and destination pairs of all the 5 flows are $s$ and $d$. 
Observe that FCS fails to assign time slots to $f_5$ after successfully scheduling the flows $f_1$, $f_2$, $f_3$, and $f_4$ as shown in Fig. \ref{fig:effects}(b2). As FCS requires all the packets of a flow to be periodically transmitted over the selected links, none of the paths have sufficient unoccupied time slots to accommodate the 2 packets of $f_5$. In contrast, after admitting the flows $f_1$, $f_2$, $f_3$, and $f_4$,  
HFS selects paths $\{s,a,d\}$ and $\{s,b,d\}$  to transmit the packets of $f_5$ as shown in Fig. \ref{fig:effects}(b3). Again, the flexibility of HFS reserving resources by scheduling each packet of flow independently brings a larger solution space of path selections as compared to FCS. 

Overall, applications that consist of flows with different cycle lengths and prioritize scalability (i.e., the number of admitted flows) would benefit from HFS as compared to FCS. On the other hand, applications that prefer periodic transmission and prioritize the minimization of delay over scalability may still use FCS rather than HFS. Consider application scenarios on TTEthernet, such as industrial control systems enabling real-time machine or robotic control, many of such scenarios can generate multiple data flows with varying cycle lengths (e.g., corresponding to different sensors' sampling rates or different controller's operational settings). Further, they require the packets to be delivered strictly within some application-specific maximum delay bound. However, as long as the delay does not exceed that bound, these applications do not require further minimization of the delay. Instead, a key optimization objective is to increase the scalability of the system by supporting a larger number of such flows using the constrained network capacity. Such application scenarios would
benefit from HFS as compared to today's FCS.

We summarize our contributions in this work as follows: 

\begin{itemize}
\item\textcolor{black}{Design of a {\it Time-Expanded Cyclic Graph (TECG)} model %
to deal with the cyclic transmission of different flows of varied cycle lengths. 
}

\item {Formulation of the hypercycle-level flexible scheduling scheme}\textemdash{HFS}. To our best knowledge, HFS is the first scheme that allows different packets of a flow to be flexibly transmitted across different cycles. 

\item The theoretical analysis for HFS's optimality.
We prove that HFS can obtain unbounded performance gain over FCS in terms of the number of admitted flows. Moreover, we prove HFS is optimal,
i.e., flexible scheduling over a larger period than a hypercycle will not outperform HFS. 

\item A load-status-based heuristic solution (HFS-LLF). We show that HFS is NP-Hard and we propose a greedy heuristic to solve HFS efficiently as a sequence of shortest-path problems. 

\item Numerical experiments to validate HFS and HFS-LLF's performance. We use real-world Ladder and AFDX topologies in our evaluation. Results show that HFS can schedule $6\times$ the number of flows scheduled by FCS. 
\textcolor{black}{Moreover, we show that HFS-LLF can achieve at least $90\%$ of the optimal result while running more than $10000\times$ faster than  solving HFS using a generic solver. 
}    
\end{itemize}

We organize this work as follows. We review the related work in Section II. In Section III, we introduce the system model and design a time-expanded cyclic graph to model available resources. We formulate the HFS problem and analyze its implementation in Section IV. We analyze the number of admitted flows under HFS vs. the optimal solutions and FCS in section V. We propose a heuristic-based algorithm in Section VI to solve HFS efficiently. 
Next, we conduct numerical experiments in Section VII. Lastly, we discuss about how to deal with varying packet size, implement HFS and apply our schemes to AoI schedule schemes in Section \ref{sec:discuss}  and summarize this work in Section \ref{sec:conclude}.

\section{Related Work}
\subsection{Works Related to Eliminating Conflicts}
In order to reduce 
resource conflicts among different flows, many studies adhere to the fixed cyclic scheduling to 
maximize the number of successfully scheduled flows without conflicts. 

For instance, Falk et al. \cite{2020RTAS_conflict} derive a conflict graph by modeling the scheduling result of each flow as a vertex and connecting two vertices with a conflict edge if the two schedules compete for a common time slot. They solve a maximum independent set problem in the conflict graph to select as many feasible schedules as possible. Zhong et al. \cite{Zhong2021} define the importance of time slots. They show that a time slot able to transmit small-cycle flow is more important than a time slot that is only able to transmit larger-cycle flow. By allocating less important slots to large-cycle flows, they avoid the conflicts of flows with different cycle lengths on important time slots. 

Moreover, Zhang et al. \cite{Zhang2022TII} use the divisibility theory to reveal under what conditions different flows will compete for resources of a time slot. With this information, they solve the maximum clique problem to accommodate the maximum number of competing flows. The work in \cite{Chen23BMSBflowfilter} also defines the compatibility between flows to filter out some flows in advance, maximizing the number of accommodated flows. The authors in \cite{xutii24conflict} further consider the computing resource consumption of source and destinations of flows. They co-optimize the deployment of source and destination of flows and the scheduling of flows, using graph partitions and group scheduling according to their proposed conflict metrics. In addition, He  et al. \cite{yaoxu'24} use the concept of time slot importance (as in \cite{Zhong2021}) to achieve a larger scheduling space than \cite{Zhong2021} through co-optimizing the path selection and scheduling.

\subsection{Works Related to Enhancing Schedulability}
Many works study how to enhance the schedulability while adopting the fixed cyclic scheduling (FCS). 
Some studies conduct joint path selection and scheduling as compared to others focusing solely on scheduling.  
Specifically, the works in \cite{huang2021online,Marek2020TCOM,yan2020injection,Zhang2022TII} study when to transmit packets along the paths that are computed or given in advance, losing the scheduling flexibility in choosing paths. To overcome this limitation,     Schweissguth et al. \cite{schweissguth2017ilp} first study the co-optimization of path selection and scheduling problems for time-triggered flows by modeling it as an Integer Linear Programming problem, which brings more schedule options. Wang et al.\cite{wang2023dmpf} seek to enhance the throughput using 
a parallel multi-path transmission scheme for time-sensitive networks. In  TTEthernet, however, each flow only has one packet to be transmitted within a cycle as in \cite{schweissguth2017ilp, Zhong2021}. Thus, the parallel transmission scheme put forth in \cite{wang2023dmpf} cannot apply to TTEthernet as one packet is the minimum transmission unit. 




Some works \cite{quan2020line,yan2020injection,tvt23xiaolongwang} have the packets proactively wait at the source nodes to achieve staggered scheduling while some schemes  \cite{huang2021online,pahlevan2019heuristic-sigbed,xutii24conflict} greedily find the earliest transmission time of flows at the source node to minimize the delay of flows. Specifically,  it is shown in  \cite{quan2020line,yan2020injection,tvt23xiaolongwang} that a network can accommodate more flows by allowing some flows to proactively wait at the source node for some time slots. References \cite{quan2020line,yan2020injection} only consider the scheduling problem assuming the path selection is done, which limits the scheduling options. To further enhance the throughput, work \cite{tvt23xiaolongwang}  considers the joint routing and scheduling problem while optimizing the best start transmission time.   


Other efforts exploit to use all nodes rather than just the source nodes to temporarily store packets before transmitting as in \cite{Marek2020TCOM,krolikowski2021computercommunications,yaoxu'24,tvt24shi}. Specifically, the authors in \cite{Marek2020TCOM} use breadth-first search to find the paths, leading to load imbalance. The authors in \cite{krolikowski2021computercommunications} model the problem as an integer linear programming (ILP) problem and solve the problem with a column generation method. More recently, 
works 
\cite{yaoxu'24} and \cite{tvt24shi}  seek to use time-varying graph-based algorithms to efficiently solve the problem for TTEthernet and Non-terrestrial networks, respectively.  The time-varying graph in \cite{tvt24shi} does not capture the cyclic feature of hypercycle while the time-varying graph in \cite{yaoxu'24} includes unnecessary edges to represent the cyclic feature of a hypercycle. 
Specifically, the time-varying graph expands the static graph in time dimension to model the status of the network in successive time periods as in \cite{peng19wcl,peng22twc,peng23infocom,tao20network,li2022enhanced}. Similar to our work here, they use a vertex to represent the node (e.g., a switch, or an end system) in a time period and add edges between vertices to represent the resource status. They also use storage edges to connect the vertices representing the same node at adjacent time periods. In this manner,  the network within different time periods are connected via the storage edges as in \cite{peng23infocom}. 
{Nevertheless,  these studies do not consider hypercycle level scheduling, which can bring significantly more flexibility in scheduling. }

\subsection{Works for Other Purposes}
Existing efforts also study the incremental scheduling problem (i.e., online scheduling). For instance, the authors in \cite{Ganesh-17-tii-incremental} model the incremental path selection and scheduling problem as ILP and are able to achieve sub-second level scheduling. Moreover, Wang et al. \cite{tii19wang} propose an enhanced scheduling scheme to rapidly respond to cluster-level and flow-level changes of train TTEthernet, by formulating a mixed integer linear programming model for offline scheduling and using a counterexample-guided methodology for online scheduling. They further propose a deep reinforcement learning scheme \cite{jiaGlobecom2021} to perform  incremental scheduling in TTEthernet.  References \cite{abdullah-19-globecom-incremental,huang2021online} incrementally identify the paths with the smallest hops and least occupation to schedule packets to be transmitted as early as possible.    

Other works such as  \cite{zhou2021reliability,Feng2022Online,Min2023Effective} study fault-tolerant time-triggered flow transmission. Yu et al.\cite{yu20adaptive} study the group routing and scheduling for multicast in time sensitive networks.  Effort \cite{sushmit22latency} studies the co-optimization of selection of source and destination nodes and the flow scheduling for better performance.  

All the above works adopt the fixed cyclic scheduling without considering the flexible scheduling method. 

\begin{table}[]
\centering
\caption{Main notations used in this paper}
\resizebox{.49\textwidth}{!}{
\begin{tabular}{r|l}
\hline
\textbf{Notations}                                                     & \textbf{Descriptions}                                                                              \\ \hline
$G,\mathcal{V},\mathcal{E}$                                   & the topology,node set and link set of TTEthernet                                          \\ \hline
$\mathcal{G},\mathcal{V'},\mathcal{E'}$                       & TECG and its vertices set and edges set                                                   \\ \hline
$\mathcal{E}'_{C},\mathcal{E}'_{S}$ & The set of  \textit{communication edges} and \textit{storage edges} in $\mathcal{E}'$. 
\\ \hline
$\mathcal{F}$                                 & the set of flows in TTEthernet, containing $N$ flows.                                                          \\ \hline
$s(i), d(i)$                                                           & a flow's source and destination node for flow $f_i$                                                     \\ \hline
$\gamma(i)$                                     & the time slot when flow $f_i$ is ready to transmit                               \\ \hline
$\lambda(i)$,$\rho(i)$                                     & flow $f_i$'s  cycle length and maximum allowed delay                               \\ \hline
$\Gamma$                                                           &  the number of time slots in a hypercycle                                           \\ \hline
$\tau_i$                                        & time slot $i$ that starts at time $t_i$ and ends at $t_{i+1}$                                    \\ \hline
$u_i$, $(u_i,v_j)$                                            & node $u$ in time slot $\tau_i$, an edge in TECG                                                   \\ \hline
$\delta^{+}(\cdot)$                                           & the set of a vertex's neighbors on outgoing edges                                         \\ \hline
$\delta^{-}(\cdot)$                                           & the set of a vertex's neighbors on incoming edges                                         \\ \hline
$\chi_k$                                                      & indicate whether flow $f_k$ is successfully scheduled                                     \\ \hline
$\mathcal{G}_{i,j}$                                           & subgraph of TECG including the topology from $\tau_i$ to $\tau_j$                          \\ \hline
$\mathcal{G}^{(k)}$                                           & the set of schedule-path graphs of flow $f_k$ in TECG                                     \\ \hline
$\mathcal{G}_{i}^{(k)}$ & $i$-th schedule-path graph of flow $f_k$                      \\ \hline
$\mathcal{V}'(k,i)$ &   the vertices set of  $\mathcal{G}_{i}^{(k)}$                    \\  \hline
$\mathcal{E}'(k,i)$ &
The edge set of $\mathcal{G}_{i}^{(k)}$
\\
\hline
$\mathcal{E}'_{C}(k,i)$ & The set of all  \textit{communication edges} in $\mathcal{E}'(k,i)$. 
\\ \hline
$\mathcal{E}'_{S}(k,i)$ & The set of  all \textit{storage edges} in $\mathcal{E}'(k,i)$ 
\\ \hline
{$x_{u,v}(k,i)$}                                              & indicates whether the $i$-th packet of flow $f_k$ traverses edge $(u,v)$\\ \hline
\textcolor{black}{$\mathcal{L}_{u,v}$ }                                  & the collection of all competing $x_{u,v}$ for the edge $(u,v)$        
\\ \hline
$\mathcal{V}'_{u}(k,i)$ & The set of representation vertices for node $u$ in $\mathcal{G}_{i}^{(k)}$

\\ \hline
$\alpha_{u,v}$                                              & the hypercycle level load status of any edge $(u_{i},v_{i+1})$ in $\mathcal{G}_{j}^{(k)}$                                     \\ \hline
$\beta_{a,b}(k,j)$& the packet level load status of an edge  $(a,b)=(u_i,v_{i+1})$ in $\mathcal{G}_{j}^{(k)}$                                    \\ \hline
$\xi_{a,b}(k,j)$                                          & the synthesized load status of an edge  $(a,b)=(u_i,v_{i+1})$ in $\mathcal{G}_{j}^{(k)}$                                 \\ \hline

\end{tabular}
}
\end{table}

\section{System Model}
\subsection{Network and Flow Model}\label{sec:flow_model}
We consider a Time-Triggered Ethernet (TTEthernet) network modeled as a graph $G=(\mathcal{V},\mathcal{E})$, where $\mathcal{V}$ is the node set,  consisting of end systems and switches and $\mathcal{E}$ is the directional links set. The nodes in TTEthernet are connected by wired full-duplex communication links. We separate each full-duplex link between node $u$ and $v$ into two directional links $(u,v)$ and $(v,u)$ and add them into $\mathcal{E}$. As such, the transmission direction of link $(u,v)$ is from $u$ to $v$. 

An end system (i.e., data source) generates a packet every cycle and transmits the packet to its targeted end system (i.e., destination) within a deterministic delay, with the relay of switches as in \cite{huang2021online,quan2020line}. We define the sequence of packets transmitted between the source and the destination systems, as a flow.  
We define a flow set  $\mathcal{F}=\{f_1,...f_i,...f_N\}$, including $N$ time-triggered flows. We denote the flow information, taken as input of the considered problem, of  flow $f_i$ by a tuple $\{s(i),d(i),\gamma(i),\lambda(i), \rho(i)\}$, where $s(i),d(i)\in\mathcal{V}$ represents the source and destination nodes, $\gamma(i)$ is the arrival time slot, i.e., the time slot in which the first packet of $f_i$ is available for transmission, 
$\lambda(i)$ is the transmission cycle's length (e.g., $\lambda(i)=4$ implies a packet of $f_i$ is sent every 4 time slots from $s(i)$) and $\rho(i)$ represents the maximum allowed delay for each packet (e.g., $\rho(i)=4$ implies that the packets need to be processed at the destination within 4 time slots from its arrival time). 
For simplicity, we represent the elements of the flow tuple as $s,d,\gamma,\lambda,\text{ and }\rho$ when $f_k$ is clear from the context. Notably, the flow information are the QoS requirements decided by the applications and taken as input, which cannot be optimized as variables.  

In particular, TTEthernet is a closed system, the information (i.e., source and destination nodes, arrival time, cycles, and delay) of all the flows is known in advance. Furthermore, we consider that all the flows in $\mathcal{F}$ appear (at possibly different time points) within the first hypercycle, and the flows do not leave/disappear in between (i.e., $\mathcal{F}$ remains unchanged). We define a hypercycle and how its length is determined in the following section. 

\subsection{Timing Definitions}

We first introduce the concept of a time slot. 
For low-complexity design of switches, we separate the considered time period into equal-length time duration\textemdash time slots as in \cite{Zhong2021,quan2020line,abdullah-19-globecom-incremental},  such that within  a  time slot $\tau$, a switch can process any packet in TTEthernet,  assuming the size of packets  are equal , and subsequently transmit it to any of its neighbor.  That is, within $\tau$, exactly one packet can cross the longest link (i.e., the link with maximum propagation delay) in TTEthernet. We also introduce how to deal with varying packet sizes in Section \ref{sec:time_slot}. 
With the concept of time slots, we measure the cycle and delay of each flow in terms of the number of time slots. Once a time slot of a link is assigned to a flow, the link is unavailable to other flows in that time slot.

Now we are ready to illustrate the concept of hypercycle.
A hypercycle $T$ specifies the minimum number of time slots within which we can maintain sufficient information of  TTEthernet resource status, such that the scheduler can schedule all the  flows. This is possible since a hypercycle is an integer times the length  of any flow's cycle. This feature allows us to flexibly schedule any flow within a hypercycle. A hypercycle $T$ includes $\Gamma$ time slots, with the $i$-th time slot $\tau_i$ starting from $t_i$ and ending at $t_{i+1}$.  

\subsection{Time-expanded Cyclic Graph}
\begin{figure}[!hbt]
\centering
\includegraphics[scale=0.9]{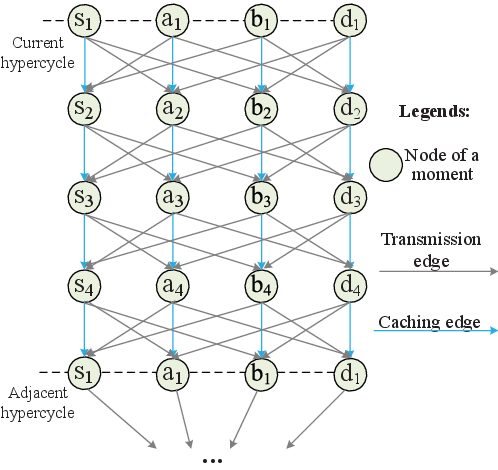}
\caption{Time-expanded Cyclic Graph.}
\label{fig:ehpg}
\end{figure}

We design a time-expanded cyclic graph~(TECG) to represent the resource status within a hypercycle $T$, similar to the time-expanded graph in previous works \cite{peng19wcl,peng22twc,peng23infocom,li2022enhanced}. Specifically, each edge in TECG belongs to a time slot, representing that the corresponding link is available in that time slot. 
Formally, we denote TECG as $\mathcal{G}=\{\mathcal{V}',\mathcal{E}'\}$, which takes as input the original topology of TTEthernet $G=\{\mathcal{V},\mathcal{E}\}$, hyper-cycle $T=\{\tau_1,...,\tau_{\Gamma}\}$ and the availability of communication link within each time slot. We illustrate the vertices and edges of TECG as follows,    
\begin{itemize}
\item \textbf{Vertex set $\mathcal{V}'$}, we generate $\Gamma$ vertices for each node $u\in \mathcal{V}$, which are denoted as $\{u_1,...u_{\Gamma}\}$, with vertex $u_i$ representing node $u$ at time {slot $\tau_i$}. 
Vertices set $\mathcal{V}'$ contains all these representation vertices. We define two functions $\delta^{-}(\cdot)$ and $\delta^{+}(\cdot)$ to identify the set of incoming neighbors and outgoing neighbors of a vertex, respectively.
For instance, function $\delta^{+}(u_i)$ represents the set of all the neighbor vertices lying on $u_i$'s outgoing edges, while $\delta^{-}(u_i)$ represents the neighbor vertices lying on $u_i$'s incoming edges. In Fig. \ref{fig:ehpg}, we have $\delta^{-}(a_4)=\{s_3, a_3, d_3\}$ and $\delta^{+}(a_4)=\{s_1, a_1,d_1\}$. 


\item \textbf{Edge set $\mathcal{E}'$}, TECG has a special structure that allows edges to only connect to vertices in adjacent time slots. 
Specifically,  let $\mathcal{E}'=\mathcal{E}'_{C}\bigcup \mathcal{E}'_{S}$, where  $\mathcal{E}'_{C}$ is a set of \textit{communication edges} and $\mathcal{E}'_{S}$ is a set of \textit{storage edges}. Each  \textit{communication edge} $(u_i,v_{i+1})$ represents a directional communication link $(u,v)$ is available in $\tau_i$. Besides, \textit{storage edges} $(u_i,u_{i+1})$  represent the process where node $u$ stores the packet within a time slot $\tau_i$. Obviously,  there is $ i\in[1,\Gamma-1]$ and $ u\in \mathcal{V}'$.
Additionally, the edges connect the vertices at time $t_{\Gamma}$ and $t_1$ to represent the resource status in $\tau_{\Gamma}$. These connections show that TECG is a cyclic graph.

\end{itemize}

We depict an example of TECG in Fig. \ref{fig:ehpg} based on a TTEthernet in Fig. \ref{fig:effects}(b) and a hypercycle $T$ of 4 time slots. The grey and blue links with arrows represent the \textit{communication edges} and \textit{storage edges}, respectively. 
We generate 4 representation vertices for each node in Fig. \ref{fig:effects}(a). Next, for each available time slot of each link, we add one corresponding \textit{communication edge} in TECG. For instance, the time slot $\tau_1$ (duration between $t_1$  and $t_2$) is available for the full-duplex link $[s,a]$, we depict edge $(s_1,a_2)$  and $(a_1,s_2)$ in TECG. Regarding \textit{storage edges}, we connect the vertices of two adjacent time slots, e.g., $(s_1,s_2)$. Notably, we add edge $(s_4,a_1)$ to represent that the link $(s,a)$ is available in $\tau_4$ and $(a_4,a_1)$ represent the storage at node $a$ within $\tau_4$. 



Notice that as defined in the edge set $\mathcal{E}'$ here we reuse the vertex names of  $s_1, \dots, d_1$ to represent the vertices in $\tau_5$ and the edges to represent the resource status in $\tau_{5}$. 
To highlight the cyclic nature of the hypercycle (where the flow assignment in one hypercycle can be repeated in every subsequent hypercycle) and from the construction of the TECG, throughout this paper,  
we consider any vertex $u_i$ to also represent the vertex 
$u_{((i-1)\%\Gamma)+1}$. Consequently, 
the vertices representing time slots for any $i$ and any $c\Gamma+i$ can be considered equivalent in the TECG, for any chosen constant $c$. {Since} we consider the index of the time slots within a hypercycle to be from 1 to $\Gamma$, we want $i$ to be $\Gamma$ if $i==\Gamma$, thus, we set $i=(i-1)\%{\Gamma}+1$ rather than directly using $i\%{\Gamma}$.


Another important thing to note is, throughout the paper, 
to make the differentiation clear between the original TTEthernet graph $G=(\mathcal{V},\mathcal{E})$ and its corresponding time-expanded cyclic graph TECG $\mathcal{G}=\{\mathcal{V}',\mathcal{E}'\}$, we call the elements in $\mathcal{V}$ nodes and  call the elements in $\mathcal{V}'$  vertices. A node of $G$ represents a switch or an end system while a vertex of $\mathcal{G}$ represents the corresponding node at a particular time slot. Similarly, we name the elements in $\mathcal{E}$ and $\mathcal{E}'$  links and edges, respectively. A link refers to the presence of a connection between two nodes, while an edge of TECG means the communication channel is available during the particular time slot. {Furthermore, we define a path as a sequence of links in graphs ${G}$, where any two adjacent links have a common node. Similarly, we define a schedule-path as a sequence of edges in $\mathcal{G}$, where any two adjacent edges have a common vertex. } A schedule-path not only gives a packet's next destination vertex 
but also the timing of the transmission 
(i.e., during which time slot).

\underline{\textit{Remark 1:}} TECG can cover the whole solution space of the resource reservation problem for flows in TTEthernet. 1):
Co-optimization of the path selection and the scheduling for flows simultaneously in a TTEthernet graph $G$ becomes equivalent to selecting schedule-paths in TECG. 
This is obvious since choosing a link to transmit a packet or choosing a node to store a packet within a time slot can be immediately mapped into an edge in TECG. 
2) TECG contains sufficient information required by resource reservation by representing the resource status of TTEthernet within a hypercycle. The resource status of TTEthernet is the same within each hypercycle. {Once the allotment for flows is done for a hypercycle, subsequent hypercycles use the same allotment.}


\section{Hypercycle-level Flexible Scheduling}


\subsection{Description of Hypercycle-level Flexible Scheduling}
We identify that the strict periodic transmission of the data packets by FCS scheduling is a sufficient (but not a necessary) solution for the problem of {cyclic packet processing} in TTEthernet. 
If the packet arrives at the destination earlier than its specified deadline, the destination node can buffer the packet according to the maximum delay experienced by the flow's packets and then pass it to the application layer for processing. 

To 
break free of this over-constraint by FCS, HFS uses individual packet level scheduling within a hypercycle to find a schedule path in the time-expanded cyclic graph for each 
packet,  from the vertex representing the source node at the packet arrival time to the vertex representing its destination node at the deadline. Thereafter, HFS reuses the identified paths in each hypercycle.



Consider a TECG
$\mathcal{G}$ and the $k^{th}$ general flow $f_k=\{s(k),d(k),\gamma(k),\lambda(k),\rho(k)\}$ as defined in Section \ref{sec:flow_model}, 
First, we define a sub-graph $\mathcal{G}_{p,q}$ of the given TECG $\mathcal{G}$ to represent TTEthernet's topology from $\tau_p$ to $\tau_q$ (including $\tau_p$ and $\tau_q$).\footnote{A TECG $\mathcal{G}$ depicts the evolved topology of a TTEthernet within a hypercycle, i.e., from $\tau_1$ to $\tau_{\Gamma}$. In regards to the sub-graph $\mathcal{G}_{p,q}$, if $q<p$, we can view $\mathcal{G}_{p,q}$ as a combination of $\mathcal{G}_{p,\Gamma}$ and $\mathcal{G}_{1,q}$. For instance, graph $\mathcal{G}_{3,1}$ of TECG in Fig. \ref{fig:ehpg} represents the vertices at $t_3$, $t_4$, $t_1$ and $t_2$,   and the edges within $\tau_3$, $\tau_4$ and $\tau_1$. Moreover, graph $\mathcal{G}_{1,1}$ includes the edges within $\tau_1$ and vertices at $t_1$ and $t_2$.}
Consider the $i^{th}$ packet of the $k^{th}$ flow $f_k$, let us denote its arrival time slot by $\gamma(k,i)$ and the {deadline} by $\bar{\gamma}(k,i) = \gamma(k,i) + \rho(k)-1$. 
For this packet, we define a schedule path graph 
$\mathcal{G}^{(k)}_i=\mathcal{G}_{\gamma,\bar{\gamma}}$, representing the TECG topology from the arrival of the $i^{th}$ packet of $f_k$ at ${\gamma}(k,i)$-th time slot to $\bar{\gamma}(k,i)$-th time slot. Similarly, given that there are no new flows entering or exiting the system, 
 a 
number of ${\Gamma}/{\lambda(k)}$ packets need to be transmitted for flow $f_k$ within a hypercycle. We denote schedule path graphs for each packet in $f_k$ by $\mathcal{G}^{(k)}=\{\mathcal{G}^{(k)}_{1},\mathcal{G}^{(k)}_{2},...,\mathcal{G}^{(k)}_{\Gamma/\lambda(k)}\}.$  We represent a \textit{schedule-path graph}   $\mathcal{G}^{(k)}_i$ {for the $i^{th}$ packet of flow $f_k$ by}  $\{\mathcal{V}'(k,i),\mathcal{E}'(k,i)\}$, where $\mathcal{V}'(k,i)$ is the vertices set and $\mathcal{E}'(k,i)$ is the edges set. Similar to $\mathcal{E}'$, we separate $\mathcal{E}'(k,i)$ into $\mathcal{E}'_{C}(k,i)$ and $\mathcal{E}'_{S}(k,i)$ to represent the set of \textit{communication edges} and \textit{storage edges}, respectively. 
We denote the path identified in the \textit{schedule-path graph}, 
from the source vertex to reach the destination vertex, by a \textit{schedule path}.

As a result, conducting HFS for a given flow set $\mathcal{F}$  equals to  finding a schedule-path in each \textit{schedule-path graph} of $\mathcal{G}^{(k)}$ for each flow $f_k$. The scheduling fails if the scheduler cannot find a schedule-path in a \textit{schedule-path graph}. Particularly, identifying a \textit{schedule-path} from its corresponding source and destination pair will naturally satisfy the maximum allowed delay requirements. Moreover,   identifying a \textit{schedule-path} in each \textit{schedule-path graph} of $\mathcal{G}^{(k)}$ will fulfill the cyclical processing requirement.

\subsection{Definitions of Variables} 


\textcolor{black}{We define a decision variable $x_{u,v}(k,i)$ for each edge\footnote{\textcolor{black}{Recall that a communication link $(u,v)$ of TTEthernet graph $G$, is represented by multiple different edges of the form $(u_i,v_{i+1})$ based on the time slot $\tau_i$. Here, we slightly abuse the notation to use $(u,v)$ to represent a TECG edge rather than a TTEthernet link.}} $(u,v)$ belonging to flow $f_k$'s $i^{th}$ packet's schedule path graph, indicating whether the packet traverses that edge or not. Specifically,  
in regards to the $i$-th packet of flow $f_k$ and its  \textit{schedule-path graph} $\mathcal{G}_{i}^{(k)}$ with an  edge set  $\mathcal{E}'(k,i)=\mathcal{E}'_{C}(k,i)\bigcup\mathcal{E}'_{S}(k,i)$,  we define a variable $x_{u,v}(k,i)$ for each edge $(u,v)$ in the set $\mathcal{E}'(k,i)$. If the packet traverses edge $(u,v)$, $x_{u,v}(k,i)=1$, otherwise, $x_{u,v}(k,i)=0$.}

\textcolor{black}{ 
Multiple packets (including packets from other flows and the same flow $f_k$) can potentially compete for the same edge of the TECG.  When the length of flow $f_k$'s maximum allowed delay is larger than its cycle length, the \textit{schedule path graphs} of different packets of $f_k$ can also overlap. Thus, the packets of the same flow can compete to transmit over the communication edge.  
Similarly, packets of different flows also compete for the same edge. For the edge $(u,v)$ we collect all such competing $x_{u,v}$
in to a set and denote that set by $\mathcal{L}_{u,v}$.The variables should satisfy the following constraints.}

\subsection{ Constraints}
To conduct a feasible HFS, each \textit{communication edge} (i.e., a time slot of a communication link) can be allocated to at most one packet, i.e., eq. (\ref{equ:path_cap_const}). Moreover, the selections of edges should formulate a schedule-path, i.e., eq. (\ref{equ:path_flow_conservation})).  There should be no loops for the selected path, i.e., a node in a schedule-path cannot be scheduled to send out a packet more than once, indicated by  eq. (\ref{equ:path_no_loop})). At last,  finding a schedule-path for each packet of a flow within a hypercycle indicates the admission of flows (see eq. (\ref{equ:path_indicator})). Specifically,   

\subsubsection{Capacity Constraints}
As mentioned above,  the packets of the same flow and of different flows can compete for the same edge. However, one \textit{communication edge}  in TECG represents an available time slot of its represented communication link, which can only be used to transmit one packet.  To enforce this constraint, 
each \textit{communication edge} $(u,v)$ can be selected at most once. There is, 
\begin{equation}\label{equ:path_cap_const}
\sum_{x_{u,v}(k,i)\in\mathcal{L}_{u,v}}{x_{u,v}(k,i)}\leq 1, \forall (u,v)\in \mathcal{E}'_C. 
\end{equation}

\subsubsection{Flow conservation}
In each \textit{schedule-path graph} $\mathcal{G}_{i}^{(k)}$ of flow $f_k$, each vertex $u$ in the vertices set $\mathcal{V}'(k,i)$, except the  source and destination vertices of the packet in $\mathcal{G}_{i}^{(k)}$,  should have same number of incoming and outgoing packets. Let $\delta^{+}(u)$ and $\delta^{-}(u)$ represent the set of neighbor vertices lying on incoming edges and the outgoing edges of vertex $u$ 
on the graph $\mathcal{G}_{i}^{(k)}$. We have, 

\begin{equation}\label{equ:path_flow_conservation}
\sum_{v\in \delta^{-}(u)}\!\!\!\!{x_{v,u}(k,i)}=\!\!\!\!\sum_{v\in \delta^{+}(u)}\!\!\!\!{x_{u,v}(k,i)}, \forall f_k\in \mathcal{F}, i\in[1,\Gamma/\lambda(k)].
\end{equation}

\subsubsection{No-loop Transmission}
A loop of a schedule-path means an entity  node of TTEthernet transmits a packet more than once. To avoid that, we can constrain each node to transmit at most one packet within a \textit{schedule path graph}. 
\textcolor{black}{For each node $u\in \mathcal{V}$ of TTEthernet, let us denote the set of all the representation vertices for $u$ in \textit{schedule-path graph} $G_i^{(k)}$ by $\mathcal{V}'_{u}(k,i)$. For instance, suppose a \textit{schedule path graph} $G_i^{(k)} = \mathcal{G}_{1,3}$ in Fig. \ref{fig:ehpg}, then $\mathcal{V}'_{a}(2,1)=\{a_1,a_2,a_3,a_4\}$. }
Moreover, the set of \textit{communication edges} in  \textit{schedule-path graph} $\mathcal{G}_{i}^{(k)}$ is $\mathcal{E}'_{C}(k,i)$. 
Thus, in each \textit{schedule-path graph}, each entity node $u$ except the destination entity node $d$ of the flow  can at most send out 1 packet, i.e., for $\forall u\in \mathcal{V}\setminus\{d\}$, we have, 
\begin{equation}\label{equ:path_no_loop}
\sum_{\{(a,b)\in \mathcal{E}'_{C}(k,i)|a\in \mathcal{V}'_{u}(k,i)\} }\!\!\!\!\!\!\!\!\!\!\!\!\!\!\!\!\!{x_{a,b}(k,i)}\leq 1, \forall f_k\in \mathcal{F}, i\in [1,\Gamma/\lambda(k)].
\end{equation}

\subsubsection{Indicator for Packets' Deadline Adherence}
For each flow $f_k\in \mathcal{F}$, we define a binary variable $\chi_k$ to indicate whether all packets of $f_k$ reach the destination before their designated deadline~($\chi_k=1$)  or not ~($\chi_k=0$). Obviously, when the solver finds a schedule path within each \textit{schedule-path graph},  scheduler successfully schedules a flow and all the flow packets reach the destination before their deadline. With the constraint (\ref{equ:path_cap_const}) and (\ref{equ:path_flow_conservation}), the scheduler successfully identifies a schedule path in $\mathcal{G}_{i}^{(k)}$ if $\sum_{v\in\delta^{+}(s_{\gamma({k,i})})}{x_{s_{\gamma({k,i})},v}(k,i)}=1$, where vertex $s_{\gamma({k,i})}$ is the source vertex of $\mathcal{G}_{i}^{(k)}$, vertices set  $\delta^{+}(s_{\gamma({k,i})})$ means the set of outgoing vertices of $s_{\gamma({k,i})}$, otherwise $\sum_{v\in\delta^{+}(s_{\gamma({k,i})})}{x_{s_{\gamma({k,i})},v}(k,i)}=0$. Thus, we have,

\begin{equation}\label{equ:path_indicator}
\chi_k \leq \sum_{v\in\delta^{+}(s_{\gamma(k,i)})}{x_{s_{\gamma(k,i)},v}(k,i)}, \forall i\in[1,\frac{\Gamma}{\lambda(k)}],f_k\in \mathcal{F}.
\end{equation}
Obviously, if scheduler fails to find one schedule-path in a schedule-path graph, $\chi_k$ can only be 0. The indicator $\chi_k$ thus can indicate whether TTEthernet admits $f_k$ or not. 

\subsection{Hypercycle-level Flexible Scheduling}
Under the above constraints, HFS aims at successfully scheduling as many flows as possible, namely,
\begin{equation}\label{equ:hfs}
\textbf{HFS:} \max\sum_{f_k\in \mathcal{F}}{\chi_k}
\end{equation}
\begin{equation}\notag
 s.t., (\ref{equ:path_cap_const})-(\ref{equ:path_indicator}).
\end{equation}
To the best of our knowledge, {HFS} is the first model which breaks the fixed cyclic scheduling routine to allow each packet of a flow to select schedule-paths more freely within a hypercycle. Moreover, {HFS} is a typical ILP problem, due to its linear constraints, objective function, and integer decision variables. {HFS 
is NP-hard as shown in the following theorem.}

\begin{theorem}
\textit{HFS is NP-hard.} 
\end{theorem}

\underline{\textit{Proof Sketch:}} 
We show the NP-Hardness of HFS {through a polynomial-time reduction from the} \textit{All-or-Nothing Multicommodity Flow}~(ANF) problem which is a known NP-Hard problem \cite{ANF-TON'24}. Specifically,  ANF is defined on a directed capacitated graph where each edge is associated with a capacity. The capacity represents the amount of data that can be transmitted through the edge. ANF considers a given set of source and destination pairs. Each pair needs to transmit a given amount of data from the source to the destination through multiple paths. ANF will earn a benefit of a positive number if a source-destination pair successfully transmits the required amount of data, otherwise, the earned benefit will be 0. ANF seeks to optimize the selections of paths for each source-destination pair to fulfill their transmission requirement without exceeding the capacity of links, to maximize the total earned benefit. 

More formally, ANF takes as input a directed graph $G$ where each directed edge $(u,v)$ is associated with a capacity $c_{u,v}$. On the graph $G$, a set of flows ${F}=\{f'_1,...,f'_N\}$ is given with flow $f'_i$  denoted by a tuple $(s'(i),d'(i),\mu(i),w(i))$, where $s'(i)$ and $d'(i)$ respectively represent the source and destination vertex of the flow $f'_i$, $\mu(i)$ denotes the amount of data to be transmitted from $s'(i)$ to $d'(i)$, $w(i)$ represents the benefits that the system can obtain by fulfilling the transmission requirement of $f'_i$. 
ANF seeks to maximize the benefits it can obtain. Now we are ready to show how to construct an HFS using ANF. Recall that HFS also takes as input a directed time-expanded cyclic graph $\mathcal{G}$ where each edge is not capacitated and a flow set $\mathcal{F}=\{f_1,...,f_N\}$ with flow $f_i$ denoted by tuple $\{s(i),d(i),\gamma(i),\lambda(i),\rho(i)\}$. The corresponding hypercycle has $\Gamma$ time slots. As such, we can first assign capacity to edges in TECG. We set $c_{u,v}=1$ if edge $(u,v)$ in TECG is a \textit{communication edge} and set $c_{u,v}=M$, with $M$ equals to the number of 
packets of $\mathcal{F}$ within a hypercycle, if $(u,v)$ is a \textit{storage edge}. \textcolor{black}{Moreover, for each 
flow in the HFS $f_i\in \mathcal{F}$, we add a virtual source vertex $s''({i})$ and add edges connecting $s''({i})$ to the source vertex of each packet in $f_i$ with a capacity of $1$. Similarly, we add a virtual destination vertex $d''({i})$ and add edges connecting the destination vertex of each packet in $f_i$ to $d''({i})$ with a capacity of $1$. }

\textcolor{black}{Let ANF take as input $G=\mathcal{G}$ and current flow set $F$. Regarding each input flow $f'_i$, we can specify it by setting $s'(i)=s''(i)$, $d'(i)=d''(i)$, $\mu(i)={\Gamma}/{\lambda(i)}$ and $w(i)=1$.  Obviously, ANF has been reduced to an HFS where delay $\rho(i)$ of each flow $f_i$ has a sufficiently large value. If there exists an optimal solver solving HFS, the admitted flow $f_i$ indicates the requirement fulfillment of flow $f'_i$ in ANF, ANF can also obtain an optimal value. As a result, HFS is also NP-hard.   }

\subsection{Formulating FCS from HFS}
We notice that adding an additional constraint to the HFS formulation results in the FCS formulation.
As compared to HFS, FCS also requires to identify a feasible schedule-path in each schedule-path graph of a flow and also needs to avoid the loop in each identified schedule-path, 
the constraints (\ref{equ:path_cap_const})-(\ref{equ:path_indicator}) thus are also needed for FCS. Moreover, FCS requires each packet of a flow to be periodically transmitted along the same path in TTEthernet. 
For  $c$ being any constant in the set of natural numbers $\mathbb{N}$, let $a= j+c\lambda$, and $b= j+c\lambda+1$, then the cyclic transmission constraint for FCS can be given as follows

\begin{equation}\label{equ:real_periodical}
x_{u_j,v_{j+1}}(k,i) = x_{u_{a},v_{b}}(k,i+c), c \in \mathbb{N}.
\end{equation}

The constraint implies that if the edge $(u_j,v_{j+1})$ was used to transmit the $i^{th}$ packet of flow $f_k$, then the edge $(u_a,u_b)$ needs to be used to transmit the $(i+c)^{th}$ packet of $f_k$ for any given value of $c\in \mathbb{N}$.
As a result, the FCS scheme can be modeled with the same objective function, i.e., eq. (\ref{equ:hfs}) and its variables subject to constraints (\ref{equ:path_cap_const})-(\ref{equ:path_indicator}) and (\ref{equ:real_periodical}). 
FCS is an ILP problem due to its integer variables, linear constraints and its objective function. {FCS is known to be an NP-hard problem as shown in \cite{schweissguth2017ilp,huang2021online}.} 

\section{Schedulability Analysis}
To rigorously study the schedulability of HFS and FCS, we first give a proof for the limitation of FCS in schedulability. Next, we prove HFS prevails over FCS in schedulability, demonstrating that the performance gain of HFS over FCS can be unbounded. Thereafter, we prove the optimality of HFS in terms of the number of flows HFS can support.

\subsection{ Schedulability Limitations for FCS}\label{sec:FCS_schedule_analysis}
Recall that, in FCS each flow $f_i$ with a cycle of length $\lambda(i)$ time slots is required to transmit a packet every $\lambda(i)$ time slots. 
We consider that the discrete time slots begin at time slot 1 and increments in steps of 1. In a particular time slot, only a single packet can be transmitted over a link. By default, we define the $\Theta_i =$  \textit{index set of $f_i$} as the sequence of time slots allocated to a flow $f_i$, i.e., $\Theta_i = \{o_i+c\cdot\lambda(i) |c\in \mathbb{N}\}$ where  $o_i$ is the index of the time slot  allocated to  the first packet of $f_i$ and $\mathbb{N}$ is the set of natural numbers. 

We identify that under FCS, two flows can be incompatible over a link under certain conditions. 

\begin{theorem}
\label{the:collapse}
 [From \textit{Theorem 1} in  \cite{Zhang2022TII}] Consider two  flows $f_i$ and $f_j$ over an unused  link $(u,v)$ with cycle lengths of $\lambda(i)$,  $\lambda(j)$, and index sets $\Theta_i$, $\Theta_j$ respectively, using FCS for scheduling. Then, there must exist a positive integer $z\in \Theta_i \cap \Theta_j$, iff $gcd(\lambda(i),\lambda(j))|(o_i-o_j)$, where $o_i,o_j$ stand for the initial indices of  $\Theta_i$ and $\Theta_j$ respectively.\footnote{Here, $gcd(a,b)$ means the greatest common divisor of integer $a$ and $b$, while $a|b$ means $b$ is dividable by $a$.}
  
\end{theorem}




We adapt the above theorem from \textit{Theorem 1} in \cite{Zhang2022TII}. We refer the interested readers to the paper \cite{Zhang2022TII} for the details of the proof. 
Using this theorem, we prove the following Lemma (and corollary) that shows, when scheduling a flow $f_i$ using FCS, the flow $f_i$ not only blocks the time slots used as given by $\Theta_i$ but also blocks some additional unused time slots for other flows $f_j$ -- indicating the poor schedulability of the FCS scheme.


\begin{lemma}
\label{lem:block_sequence}
\textit{Under FCS,  allocating time slots  $\Theta_i = \{o_i+c_i\cdot\lambda(i) |c_i\in \mathbb{N}\}$  of an unused link $(u,v)$ to  flow $f_i$  with cycle $\lambda(i)$ will block other flows $f_j$ with cycle $\lambda(j)$  from using (unused) time slots $\{o_i+c\cdot\lambda|c\in \mathbb{N}\}$  with  $\lambda=gcd(\lambda(i),\lambda(j))$. }   
\end{lemma}

\noindent\underline{\textit{Proof Sketch:}}
This can be proved by contradiction. Say FCS does not block any additional time slots and the time slot $o_i+c\cdot\lambda$ (for  $\lambda=gcd(\lambda(i),\lambda(j))$ and some $c\in \mathbb{N}$) 
is available and allocated to flow $f_j$. 
It implies that $o_i+c\cdot\lambda=o_j+c'\cdot\lambda(j)$, for some positive integer $c'$.  Moving terms of this equation, we have $o_j-o_i=c'\cdot\lambda(j)-c\cdot\lambda$. It follows from this equation that the $gcd(\lambda(i),\lambda(j))$ will be divisible by $(o_j-o_i)$, i.e., $gcd(\lambda(i),\lambda(j))|(o_j-o_i)$. 
It implies from Theorem \ref{the:collapse} that there must exist some integer $z$ where $f_i$ and $f_j$ collide, resulting in a contradiction, and thereby proving the Lemma.

\begin{corollary}
\label{corol:co-prime}
\textit{The index sets $\Theta_i$ and $\Theta_j$ of flows $f_i$ and $f_j$ over an unused link $(u,v)$ determined through FCS scheduling, will always collide regardless of the initial indices $o_i$ and $o_j$, if the cycle length of the flows, $\lambda(i)$ and $\lambda(j)$ are co-prime.} 
\end{corollary}

\noindent\textit{\underline{Proof Sketch}:} 
This corollary implies that using FCS scheduling, after time slots are allocated to a flow $f_i$ with cycle $\lambda(i)$ over link $(u,v)$, any flow $f_j$ that has a co-prime cycle with $\lambda(i)$ (i.e., $gcd(\lambda(i),\lambda(j))=1$) cannot be admitted by link $(u,v)$ even if link $(u,v)$ still has many (unused) available time slots. Since the cycles $\lambda(i)$ and $\lambda(j)$ are co-prime, there is  $\lambda=gcd(\lambda(i),\lambda(j))=1$. \textit{Lemma \ref{lem:block_sequence}} immediately indicates that the accommodation of flow $f_i$ will block $f_j$ from using any time slots $\{o_i+c|c\in \mathbb{N}\}$. This further indicates two index sets will always collide.

An interesting observation from Lemma \ref{lem:block_sequence} is that,  once allocating time slots $\gamma(i)$ to a flow $f_i$ with a cycle $\lambda(i)$, the time slots sequence  $\{o_i+c\cdot\lambda|c\in \mathbb{N}\}$ (containing both used and unused time slots by flow $f_i$) becomes permanently unavailable (or blocked) for another flow $f_j$ with cycle length $\lambda(j)$, where $\lambda=gcd(\lambda(i),\lambda(j))$. The smaller $\lambda$ is, the denser the blocked set $\{o_i+c\cdot\lambda|c\in \mathbb{N}\}$ becomes. 
Corollary $\ref{corol:co-prime}$ describes the extreme condition when $\lambda=1$ ($\lambda(i)$ and $\lambda(j)$ are co-prime), where all the unused time slots of $f_i$ becomes unavailable to flow $f_j$. As a by-product,  parameter $\lambda$ can be used to evaluate the degree of incompatibility of two flows with cycles $\lambda(i)$ and $\lambda(j)$, scheduled under FCS. The smaller $\lambda$ is, the more incompatible the two flows can be over an unused link or path.

\subsection{Schedulability Analysis for HFS vs. FCS}\label{sec:HFS_schedule_analysis}
Given FCS schedulability limitations as demonstrated in the previous section, here we show that HFS significantly outperforms FCS in terms of schedulability. It does so by getting rid of the incompatibility between flows, resulting in a larger solution space of feasible schedules than FCS.

\begin{lemma}\label{lemma:larger_space}
\textit{HFS has a larger solution space of feasible solutions than that of FCS. }
\end{lemma}

\textit{\underline{Proof:}} We can prove the lemma by showing that all valid FCS schedule are also valid HFS scheule, but not all valid HFS schedules are FCS schedules, i.e., the feasible solutions of FCS satisfy the constraints in HFS, while the feasible solutions of HFS do not necessarily satisfy the constraints of FCS.  Let us consider a general feasible solution of FCS, satisfying constraints of FCS including (\ref{equ:path_cap_const})-(\ref{equ:path_indicator}) and (\ref{equ:real_periodical}), clearly, it also satisfies the constraints of HFS, i.e.,   (\ref{equ:path_cap_const})-(\ref{equ:path_indicator}). However,  the feasible solution of HFS satisfying  (\ref{equ:path_cap_const})-(\ref{equ:path_indicator}) might violate constraint  (\ref{equ:real_periodical}), which indicates the solution of HFS may not necessarily be the solution of FCS. Hence, HFS has a larger solution space of feasible solutions than that of FCS.


The following lemma demonstrates the better schedulability of different flows in HFS as compared to FCS. 

\begin{lemma}\label{prop:compatibility}
\textit{When scheduling two flows $f_i$ and $f_j$ over an unused link $(u,v)$ using HFS, allocating time slots to $f_i$ over link $(u,v)$ will not block $f_j$ from using the remaining unused/unoccupied time slots. } 
\end{lemma}

\textit{\underline{Proof:}} Under FCS, constraints (\ref{equ:path_cap_const}) and (\ref{equ:real_periodical}) incur competition among different flows for time slots. The constraint (\ref{equ:path_cap_const}) ensures that the used/occupied time slots by $f_i$ should not be available to other flows, while constraint (\ref{equ:real_periodical})  enforces the cyclic occupation of time slots on the link. In fact, the constraint (\ref{equ:real_periodical}) is the premise of the establishment of \textit{Theorem \ref{the:collapse}} 
and \textit{Lemma \ref{lem:block_sequence}} that highlight the restrictive scheduling of FCS. Therefore, after HFS removes the constraint (\ref{equ:real_periodical}),  
the otherwise unused yet blocked time slots become available for allocation to any other flow $f_j$.

\begin{theorem}
\textit{ The performance gain, measured by the number of flows that can be admitted by HFS vs. FCS, can be unbounded.}
\end{theorem}

\textit{\underline{Proof:}} 
We prove the theorem in two parts. In the first part, we prove that a solution obtained through HFS scheduling is at least as good as one obtained through FCS scheduling. This immediately follows from Lemma \ref{lemma:larger_space} and the fact that FCS has additional constraints than HFS. 

In the second part of the proof, we show a network construction where HFS admits $n$ times more flows as opposed to FCS, for any chosen positive (integer) value of $n$. Let us consider a line network and a set of flows to be transmitted from one end to the other end over the line network. Considering a set of cyclic flows  whose cycles are a set of co-prime numbers $\{p_1, .... p_n\}$, such that $\sum_i 1/p_i < 1$ (i.e., the $n$ prime numbers are all greater than $n$) over a link $(u,v)$ with $n$ unallocated time slots. Suppose the delay of each flow is the same or greater than its cycle length. For all links in the line network, according to \textit{Corollary \ref{corol:co-prime}}, FCS can admit only 1 flow since the cycle lengths of any two flows are co-prime. Given HFS's flexibility and the fact that the allowed delay being greater than the cycle length, it is easy to see that all $n$ flows can be admitted using HFS. Since we can create this for any chosen $n$, the performance gain of HFS measured in terms of the number of flows as compared to FCS can be unbounded.

\subsection{Optimality Analysis of HFS}

In this section, we ask the question: Does a more flexible scheduling scheme than HFS exist? and, will such a flexible scheduling (over a larger time range than the hypercycle) outperform HFS? We identify the answer as ``no''.  
Intuitively, the most flexible scheduling would be a packet-level flexible scheduling (PFS), i.e., flexibly scheduling each packet of a flow within the flow's lifespan (i.e., from when the flow arrives to when the flow exists TTEthernet). PFS considers all possible packet assignment possibilities and chooses the one that maximizes the number of flows admitted.  

\begin{observation}\label{obv:pfs}
\textit{The number of flows admitted by the PFS scheme is optimal.}      
\end{observation}

We {illustrate the superiority} 
of HFS in the following theorem by showing that it can achieve the same performance as that of the most flexible PFS scheme. In comparison to PFS, HFS however requires much less computation capability.

\begin{theorem}
\textit{Given an arbitrary set of flows  whose maximum allowed delays have lengths no smaller than the lengths of their cycles and no greater than the length of a hypercycle
, HFS can admit the same set of flows as PFS does over a link.} 
\end{theorem}

\noindent\textit{\underline{Proof:}} Recall that we assume a static set of flows, i.e., all the flows start from the first hypercycle. From the second hypercycle onwards, the allocation decision under HFS must remain the same across the hypercycles.

We prove the theorem through contradiction.
Say there is a schedule in PFS that admits more flow than HFS.
Then, 
1) there must exist at least one hypercycle with no total resource deficit, i.e., $\sum_i c_i' \geq \sum_i c_i$, where $c_i$ is the requested time slots for flow $f_i$, and $c_i'$ is the allocated slots under this particular hypercycle. This is true, because otherwise, PFS cannot sustain the flows.
2) In this hypercycle, if for all $f_i$, $c_i' \geq c_i$, then the HFS allocation is done by just repeating this exact hypercycle resource allocation, leading to a contradiction.
3) Otherwise, in the hypercycle, if there is some flow $f_i$, for which $c_i' < c_i$, then there must exist other flow $f_j$  with $c_j'> c_j$ (as $\sum_i c_i' \geq \sum_i c_i$). Since our considered flows have their maximum allowed delays no smaller than the lengths of their cycles and no greater than the length of a hypercycle, 
this mismatch between flows in a hypercycle can only happen when the arrival time and deadline of a flow's 
packet 
lie
in two adjacent hypercycles. 
For instance, suppose link $(u,v)$ admits a flow $f_k$ with a cycle length of 2 time slots and delay requirement of 2 time slots under PFS. Let the hypercycle be of 10 time slots and the first packet of $f_k$ arrive at $\tau_2$ (the 2nd time slot). In the second hypercycle, PFS can achieve $c_k’>c_k$ only by allocating both $\tau_{11}$ and $\tau_{20}$ (the $11$th and $20$th time slots) to $f_k$. That is, flow $f_k$'s last arrived packet of first hypercycle and last arrived packet of the second hypercycle are both transmitted within the second hypercycle. Similarly, to achieve $c_k'<c_k$ in second hypercycle, the only way is to transmit $f_k$'s last arrived packet of first hypercycle in the first hypercycle but to also transmit the last arrived packet of second hypercycle in the third hypercycle. With this understanding, for any $i$ and $j$ in the considered hypercycle with $c_i'>c_i$ and $c_j'<c_j$, we can directly retrieve the time slots allocated to the repeatedly transmitted packets of $f_i$ and reallocate them to new time slots of $c_j$. After the reallocation, we have all $i$, satisfying $c_i' \geq c_i$, then the previous argument applies.

The following corollary immediately follows from the above theorem and observation \ref{obv:pfs}, which shows HFS's optimality for a large number of cases with bounded allowed delay.

\begin{corollary}
\textit{Given an arbitrary set of flows  whose maximum allowed delays have lengths no smaller than the lengths of their cycles and no greater than the length of a hypercycle, the number of flows admitted by HFS is optimal.} 
\end{corollary}

\section{Efficient Graph-based Algorithm}
Due to the NP-Hardness of HFS, we put forth a graph-based scheme to efficiently solve HFS in an incremental manner. This scheme can also be applied in an online environment. Specifically, 
we first design a metric to represent the hypercycle-level and packet-level load status of each link. Next, based on that,  we design a lightest-load first hypercycle-level flexible scheduling algorithm (HFS-LLF) that aims to find proper schedule-paths for each flow such that the network can accommodate more flows.     

\subsection{Hypercycle-level and Packet-level Load Status of links}
To maximize the number of scheduled flows, the algorithm should: 1) identify feasible schedule-paths for each flow, i.e., schedule-paths that satisfy the constraints in HFS. 2) minimize the effect of allocating resources to the current flow on resource allocation for other flows, as different flows share the resources as indicated in the constraint 
(\ref{equ:path_cap_const}). 
One effective way is to design a load-balance flow allocation algorithm to minimize the effect. 


Given the high schedulability of HFS, it can support flexible allocation of the time slots of links, suitable for achieving a load-balanced allocation.
Given an TECG  within a hypercycle $T$, we define two levels of load status for each edge. For a link $(u,v)$, its 
hypercycle-level load status determines the availability of unused time slots within a hypercycle, and 
is given by the fraction of time slots (within $T$) where a link is being used. Namely, for any link  $(u,v)$ within a hypercycle, we have,  




\begin{equation}\label{equ:space_load}
\alpha_{u,v} = \frac{|(u,v) \text{\textit{'s occupied slots within a hypercycle}}|}{\Gamma}, 
\end{equation}

Here, 
we can easily check the number of unoccupied time slots of link $(u,v)$ within a hypercycle by counting the number of edges in the TECG representing the link $(u,v)$ as each edge in TECG represents an available and unoccupied time slot of a link. Next, we can subtract the count from $\Gamma$ to obtain the number of occupied time slots. 
Each representation edge of the link  $(u,v)$ has the same value of $\alpha_{u,v}$. 

Notice that, load-balance only at the hyper-cycle level might not be sufficient. For instance, a lightly loaded link (at the hypercycle level) can still have some consecutive occupied time slots. This is not desired since other flows may need a time slot from the time duration of the consecutively occupied time slots to satisfy its cyclic transmission requirement. 


For a more fine-grained load balance, we will evaluate the load status of a link, when selecting paths for a packet, within a packet's lifespan (from when the packet arrives to the packet's deadline). We call such load status as the packet level load status. Specifically, when selecting the path for the $i$-th packet of flow $f_k$ from edges set $\mathcal{E}'(k,i)$,   for each edge $(a,b)\in \mathcal{E}'_{C}(k,i)$ representing link $(u,v)$,\footnote{Recall that a link $(u,v)$ is represented by multiple different edges of the form $(u_j,v_{j+1})$ on the TECG based on the time slot $j$.} we have its packet level load status for the $i^{th}$ packet as,

\begin{equation}\label{equ:time_load}
\beta_{a,b}(k,i)=
 \begin{cases}
       	\frac{|(u,v) \text{\textit{'s occupied slots within the lifespan}}|}{\rho(k)} & \text{if}\ \rho(k)< \Gamma; \\
      	 \alpha_{u,v} & \text{if}\ \rho(k)\geq \Gamma. \\
      	
    \end{cases} 
\end{equation}

Similarly, we can count the number of edges in set $\mathcal{E}'_C(k,i)$ representing link $(u,v)$ 
 to obtain the number of $(u,v)$'s unoccupied time slots within the lifespan of the packet. We can then subtract the count from $\rho(k)$ to obtain the number of occupied time slots. 
 The scheduler can select edges with a smaller $\beta$ when selecting paths for a packet to avoid occupying a link in consecutive time slots.

The combined load status for each edge $(a,b)$ representing link $(u,v)$ when selecting edges in TECG for $i$-th packet of flow $f_k$ is the sum of its hypercycle-level and packet-level load status, namely, 
\begin{equation}\label{equ:congestion_status}
\xi_{a,b}(k,i)=
\alpha_{u,v}+\beta_{a,b}(k,i), \forall (a,b)\in \mathcal{E}'_{C}(k,i).
\end{equation}
 In regards to the storage edges, since their capacity are sufficient, we set their load status to be 0.

With the representation of hypercycle-level and packet-level load status for each edge, we can heuristically use shortest path algorithms to  identify the schedule path with lightest load with the aim to accommodate as many flows as possible. 

\subsection{Load Balanced Pathfinding Algorithm}
In this subsection, we put forth a pathfinding algorithm in TECG to achieve balanced load while maintaining the feasibility.    

As shown in Algorithm \ref{alg:small-load-algorithm}, the main idea is to find a schedule path with lightest load  for each packet. 
Specifically, before starting to find the schedule-path for the $i$-th packet of flow $f_k$,  we first obtain the arrival time slot $\gamma_{k,i}$ and deadline $\bar{\gamma}_{k,i}$ and update the edges weight of load status in the \textit{schedule-path graph} (see line (6-7)). Next, we temporarily remove the edges of the  $(\bar{\gamma}_{k,i}+1)-$th time slot to disconnect the graph and potentially prevent the algorithm from obtaining looped paths. We store the removed edges in an edge set $\mathcal{E}_t$ (see line (8-9)). Since the graph $\mathcal{G}$ is a cyclic graph, a packet can potentially loop infinitely, 
e.g., as in Fig. \ref{fig:ehpg}, suppose a packet arrives at $s_2$ and needs to reach $d_4$, this packet can reach $d_4$ via traversing a sequence vertices say $(s_2, b_3, d_4)$, or $(s_2,a_3,a_4,a_1,a_2,d_3,d_4)$, and so on. Removing the edges of $(\bar{\gamma}_{k,i}+1)-$th time slot temporarily disconnects the graph and helps the pathfinding algorithm to only obtain paths that satisfy the given delay requirements. 

After temporarily removing the edges, the algorithm finds a schedule-path with the smallest load. If there exists no \textit{schedule-paths}, the algorithm fails to ensure periodic transmission. The algorithm restores all the previously removed edges in $\mathcal{P}$ and $\mathcal{E}_t$ to $\mathcal{E}'$, and then stops with the output of an empty set. Otherwise, the algorithm records the identified schedule-path in $\mathcal{P}$ and removes the communication edges in the identified path from graph $\mathcal{G}$, representing the fact that they become occupied (see line (9-17)). After identifying a schedule path, the algorithm restores the removed edges from $\mathcal{E}'_{t}$ to $\mathcal{E}'$ (see line 18) for identifying a path in next \textit{schedule-path graph}.   

\begin{algorithm}[th]\caption{Lightest-load First Hypercycle-level Flexible Scheduling (HFS-LLF) Algorithm }\label{alg:small-load-algorithm}
\begin{algorithmic}[1]
\STATE {\textbf{Input:} Flow {$f_k=\{s,d,\gamma,\lambda,\rho\}$,} time-expanded cyclic graph $\mathcal{G}=\{\mathcal{V}',\mathcal{E}'\}$.}
\STATE{\textbf{Output:} Schedule-paths set $\mathcal{P}$ for each packet.}
\STATE{\textbf{define} an edge set $\mathcal{E}_{t}$ to temporarily store edges.}
\FOR{$1\leq i \leq \Gamma/\lambda$}
\STATE{\textbf{set} $\mathcal{E}_{t}=\varnothing$.}
\STATE{\textbf{obtain}  arrival time and deadline $\gamma(k,i)$ and $\bar{\gamma}(k,i)$.}
\STATE{\textbf{update}  weights of edges in $\mathcal{E}'_{C}(k,i)$ using Eq. (\ref{equ:congestion_status}).}
\STATE{\textbf{add} the edges in $(\bar{\gamma}(k,i)+1)$-th time slot in $\mathcal{E}'$ into $\mathcal{E}_t$ and temporarily remove them from $\mathcal{E}'$.}
\IF{ there is a path \textcolor{black}{from $s_{{\gamma}(k,i)}$ to $d_{\bar{\gamma}(k,i)+1}$}}
\STATE{\textbf{identify} the shortest path w.r.t. edge weight $\xi$ via shortest path algorithm\cite{complexity-dijsktra}.}
\STATE{\textbf{add} the identified path to $\mathcal{P}$.}
\STATE{\textbf{remove} the edge of the path belonging to $\mathcal{E}'_{C}(k,i)$ from $\mathcal{E}'$.}
\ELSE
\STATE{\textbf{restore} each edge in $\mathcal{P}$ to $\mathcal{E}'$.}
\STATE{\textbf{restore} all edges in $\mathcal{E}_t$ into $\mathcal{E}'$.}
\STATE{\textbf{terminate} the algorithm with $\mathcal{P}=\varnothing$.}
\ENDIF
\STATE{\textbf{restore} all edges in $\mathcal{E}_t$ into $\mathcal{E}'$.}
\ENDFOR
\end{algorithmic}
\end{algorithm}

\subsection{Feasibility Analysis For HFS-LLF}
\begin{proposition}
The schedule-paths identified by algorithm HFS-LLF satisfy capacity constraint (\ref{equ:path_cap_const}). 
\end{proposition}

\underline{\textit{Proof:}} Every time we find an identified-path, we will remove the transmission edges in $\mathcal{E}'$ as indicated in line (12). Thus, a \textit{transmission edge} will never be occupied more than once, satisfying constraint   (\ref{equ:path_cap_const}) . 

\begin{proposition}
The schedule-paths identified by algorithm HFS-LLF satisfy flow conservation constraint (\ref{equ:path_flow_conservation}). 
\end{proposition}

\underline{\textit{Proof Sketch:}} The shortest path algorithm  naturally requires every traversed vertex except the source and destination vertex to have one selected incoming edge and one selected outgoing edge along the path, satisfying constraint (\ref{equ:path_flow_conservation}).  

\begin{proposition}
The schedule-paths identified by algorithm HFS-LLF satisfy no-loop constraint (\ref{equ:path_no_loop}). 
\end{proposition}

\underline{\textit{Proof Sketch:}} We use the proof by contradiction to prove this \textit{proposition}. Suppose the algorithm identifies a shortest schedule-path including a loop $\{a_m,b_i,...,c_j,a_n\}$ with $b\neq a$ and $c\neq a$. This loop contains at least two \textit{transmission edges}, thus the total load status weight of each edge in this loop is positive according to Eq. (\ref{equ:congestion_status}). Meanwhile,  there always exists a  schedule-path $p_1$ from $a_m$ to $a_n$ in TECG, where $p_1=\{a_m,a_{m+1},...,a_{n-1},a_{n}\}$ comprising only of successive \textit{storage edges}. The sum of the weights $\xi$ of this storage schedule-path $p_1$ is 0 since the synthesized load status of each \textit{storage edge} is 0. As a result, there will exist a shorter path than the identified schedule-path via replacing the loop $\{a_m,b_i,...,c_j,a_n\}$ in the identified scheduled path with $p_1$ resulting in a contradiction. Therefore, the identified schedule-path always satisfy constraint (\ref{equ:path_no_loop}).   

\begin{proposition}
The schedule-paths identified by algorithm HFS-LLF satisfy the deadline adherence constraint (\ref{equ:path_indicator}). 
\end{proposition}

\underline{\textit{Proof Sketch:}} We prove this by contradiction. Recall that constraint (\ref{equ:path_indicator}) implies a flow is successfully admitted under HFS if each packet of a flow within a hypercycle is assigned a schedule path.  Say algorithm \ref{alg:small-load-algorithm} admits a flow with at least one packet not assigned a schedule path. Algorithm 1, however, stops  with no output once no paths are found for a packet according to step 13-17, leading to a contradiction. 
Generally, algorithm \ref{alg:small-load-algorithm} will only admit a flow $f_k$ and output the corresponding paths  $\mathcal{P}$ when it identifies a schedule-path for each packet of a flow. These operations clearly align with what constraint (\ref{equ:path_indicator}) indicates. 
As a result, the identified schedule-paths satisfy all the constraints of HFS, indicating their feasibility. 

\subsection{Complexity Analysis}
\begin{lemma}
The worst-case time complexity of Algorithm \ref{alg:small-load-algorithm} is $O(\frac{\Gamma}{\lambda}\rho(|\mathcal{E}|+|\mathcal{V}|\log{|\mathcal{V}|}))$, where $\Gamma$ is the number of time slots within a hypercycle, $\lambda$ is the flow's cycle, $\rho$ is the flow's delay requirements, $|\mathcal{E}|$ is the number of directed links  and $|\mathcal{V}|$ is the number of nodes in TTEthernet.  
\end{lemma}

\underline{\textit{Proof:}} 
The main operation for Algorithm \ref{alg:small-load-algorithm} is to find $\frac{\Gamma}{\lambda}$ schedule-paths. To find a schedule-path, the time complexity of updating load status weight of edges is $O(\rho|\mathcal{E}|)$,  the time complexity of removing and adding edges is $O(|\mathcal{E}|)$, and the time complexity of finding a shortest schedule-path  within $\rho$ time slots is $O(\rho|\mathcal{E}|+\rho|\mathcal{V}|\log{\rho|\mathcal{V}|})=O(\rho(|\mathcal{E}|+|\mathcal{V}|\log{|\mathcal{V}|))}$ using a Fibonacci heap as in \cite{complexity-dijsktra}.
Thus, the time complexity for finding a schedule-path is $O(\rho|\mathcal{E}|+|\mathcal{E}|+\rho(|\mathcal{E}|+|\mathcal{V}|\log{|\mathcal{V}|)}=O(\rho(|\mathcal{E}|+|\mathcal{V}|\log{|\mathcal{V}|))}$. Finally, the total complexity for finding $\frac{\Gamma}{\lambda}$ schedule-paths, i.e., $O(\frac{\Gamma}{\lambda}\rho(|\mathcal{E}|+|\mathcal{V}|\log{|\mathcal{V}|}))$. 

\subsection{Masking the jitter introduced by HFS}
As HFS/HFS-LLF flexibly schedules each packet in a hypercycle, it can lead to non-periodic arrival times at the destination nodes at the network layer. To solve this issue, HFS/HFS-LLF let the destination node buffer each received packet within a flow according to the largest delay experienced by the flow's packets, rather than immediately passing it to the application layer of the destination node upon receipt. This approach ensures the application layer of the destination node can receive the packets in a periodic manner. In particular, buffering the data at the destination node to smooth the data arrival pattern is a common technique that many works (e.g., \cite{buffer_2014}) exploited 
to overcome such jitter. Such a buffering mechanism is at the cost of increased average packet delay for HFS, but recall that under HFS, even the packet with the largest delay still satisfies the flow's deadline requirement. We will use a concrete example to illustrate how this can be implemented in Fig. \ref{fig:example_implementation}  later (see Section \ref{sub:implementation_hfs}).

\section{Experiment Results}
In this section, we first illustrate the simulation setup and analyze the evaluation results in terms of the running time, finished number of flows, and the throughput. 

\subsection{Experiment Setup}
We run all the experiments in an Intel(R) Four-Core (TM)  i5-1135G7 CPU with 2.4 GHz and 16 GB of RAM. We implement our proposed \textit{HFS} scheme by inserting the objective function (\ref{equ:hfs}) along with its constraints into mathematical solver Gurobi 11.0.1\cite{Gurobi}. Besides, we implement algorithm \textit{HFS-LLF} using the package Networkx in Python.

\subsubsection{Evaluated Schemes} We introduce the baseline schemes used for the evaluation and how they are implemented as follows.  
\begin{itemize}
\item \textit{FCS}, as proposed in existing joint path selection and scheduling works,  cyclically transmits each packet of a flow, while allowing temporary storage at each node as in \cite{Marek2020TCOM,krolikowski2021computercommunications,yaoxu'24,tvt24shi}. We implement the scheme using a mathematical solver Gurobi to enforce the constraints and objective functions.
\item\textit{JRAS-TSEG}, as proposed in \cite{yaoxu'24}, jointly determines when and which link to transmit data, allowing data to be temporarily stored on nodes. We implement this scheme using shortest path algorithm provided by Networkx package\cite{Networkx} in Python. 
\item \textit{BFS-S}, obtains the path by a breadth-first search 
(BFS) method and allows the node to temporarily store the data as 
 in \cite{Marek2020TCOM}. We implement this scheme by applying the BFS algorithm in our TECG. 
 
\item \textit{IRAS}, as proposed in \cite{huang2021online}, selects all the feasible paths for one packet and then sorts the paths according to the number of hops and the number of flows assigned on the paths. Then IRAS tries each sorted path to schedule the current flow. In particular, IRAS is a no-wait scheme that does not allow the packets to be temporarily stored at nodes. 
We implement this scheme using the package\textemdash Networkx\cite{Networkx} in Python.
\end{itemize}

\subsubsection{Topology of Experiment Network}
Similar to work\cite{Zhong2021}, we evaluate the above schemes under 3 networks: 1) AFDX (Avionics Full-Duplex
switched Ethernet) for airbus A380, known for providing real-time data delivery.  The AFDX network comprises 9 nodes and 13 links; 2) LADDER topology for trains communication network. LADDER network is introduced in IEC 61375-3-4 (Ethernet Consist Network), 
which is an international standard of train communication network as in \cite{Zhong2021}. The LADDER network consists of 8 nodes and 9 links; 3) a randomly generated network with 50 nodes represented by the erdos rényi random graph as in \cite{erdos1960evolution}. A link is established between any two nodes with a probability $p$. 

\subsubsection{Settings for Networks and Flows}
For each considered network topology, we configure each node with a sufficiently large buffer size. The data transmission rate for each communication link is set as 1 Gigabit per second (Gbps) as in \cite{huang2021online,Zhang2022TII,quan2020line}. Besides, the length of a time slot is set to \SI{15}{\micro\second} to allow any packet to cross a communication link. We generate the flows referring to IEC/IEEE 60802 standard\cite{60802,yan2020injection,huang2021online} which describes the flow features in different actual industrial scenarios. By default, we randomly choose the source and destination nodes for flows from the node set and remove the duplicate flows that have the same source and destination. 
Besides, we set the length of each flow's allowed delay the same as the length of the flow's cycle. \textcolor{black}{The cycles of flows in different experiments are different. We specify them in each experiment.}

\subsection{Performance Gain of  HFS over FCS}

\subsubsection{Gain in Scheduling}
We first consider a setting 
where the gain of throughput only reflects the gain obtained purely from the scheduling differences by ensuring that the flow path remains the same in the different schemes.  Given a link, scheduling decides using which time slot of this link to transmit which flow's packet. To explore the gain in scheduling, we assume each flow's path is of only 1 hop, i.e., a flow's source and destination nodes lie in the same link. As such, in AFDX (containing 28 directed links) and LADDER (containing 20 directed links) topologies, let the number of arriving flows for each link increase from 1 to 6, i.e., the total number of flows increases with a step of 28 in AFDX and 20 in LADDER. The cycle lengths of the flows on each link are  3, 5, 7, 11, 13, and 17, respectively. In such a manner, HFS and FCS will always choose the same path for a flow and the gain in throughput completely originates from the scheduling.  

\begin{figure}[htbp]
    \centering
    \subfigure[AFDX]{
    \centering
    \includegraphics[scale=0.23]{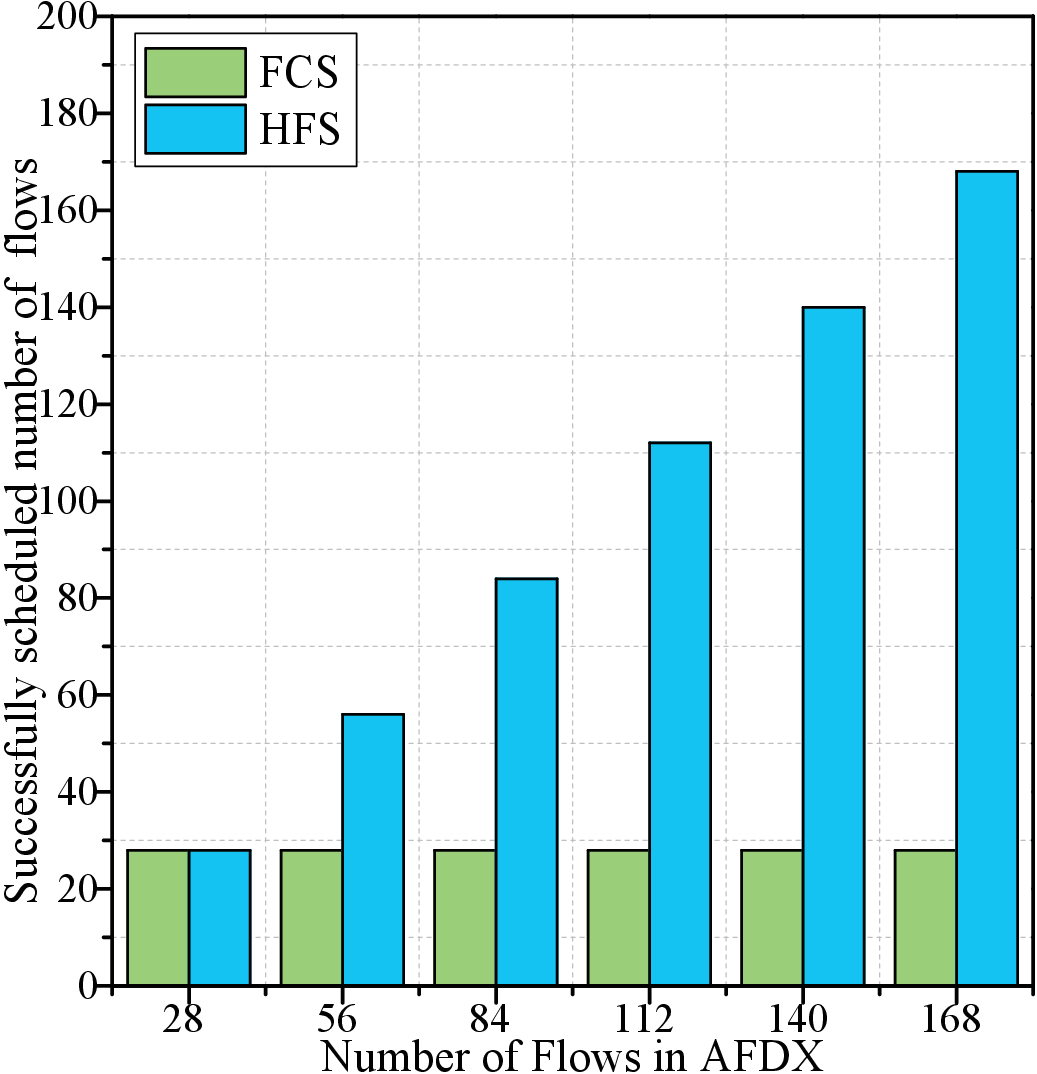}
    }
    \subfigure[Ladder]{ 
    \centering
    \includegraphics[scale=0.23]{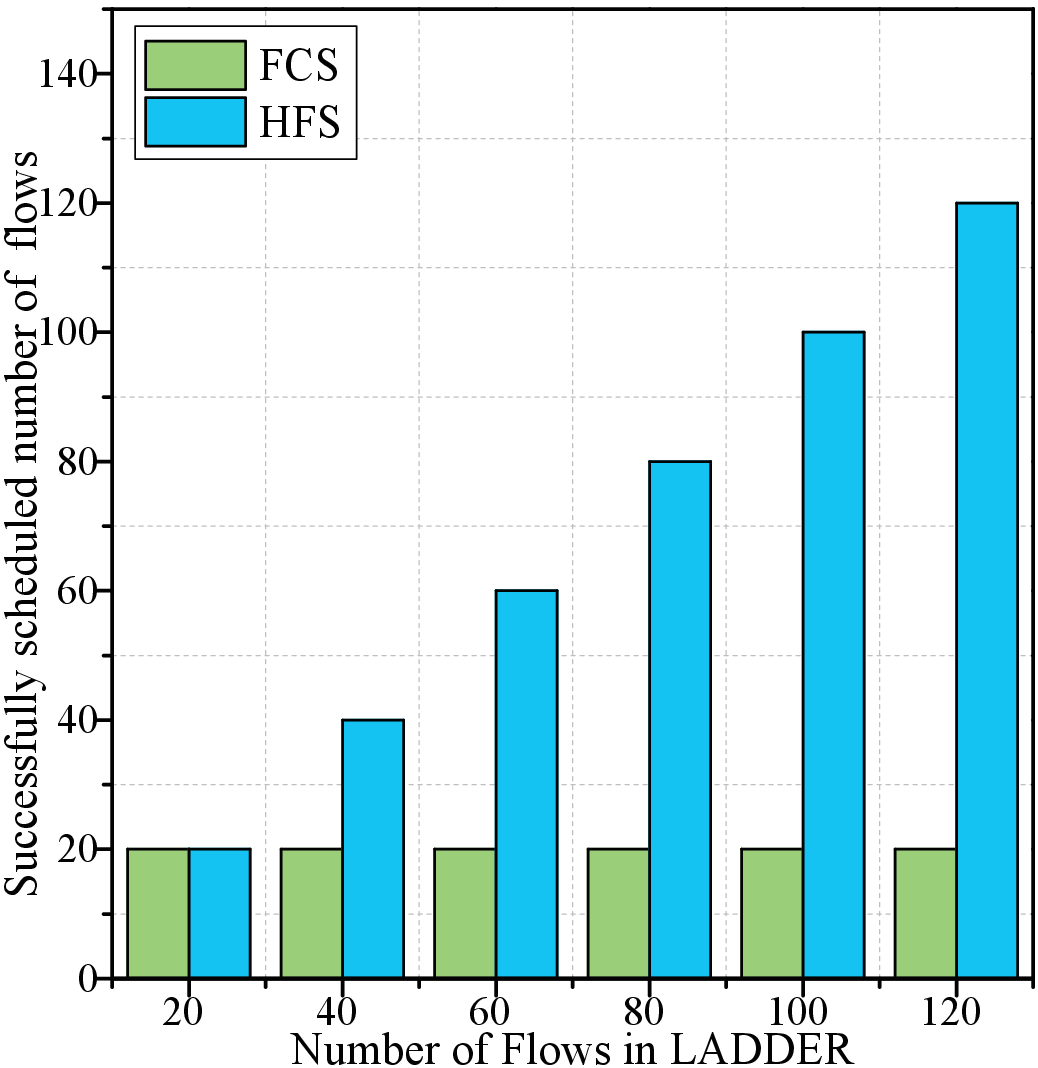}
    }
    \caption{Number of successfully scheduled flows by FCS and HFS schemes on AFDX and Ladder topologies.}    \label{fig:scheduling_gain}
\end{figure}

As the number of inserted flows increases, Fig. \ref{fig:scheduling_gain}(a) and Fig. \ref{fig:scheduling_gain}(b) plots the number of successfully scheduled flows in AFDX and Ladder topology, respectively. As more flows join in, both in AFDX and Ladder topology, FCS is unable to schedule those flows while HFS successfully schedules each of the newly arrived flows. This is as expected. Specifically, after FCS chooses a time slot to transmit one packet of a flow, it determines the choice of the time slots to transmit other packets of the flow as well. In contrast, within a hypercycle, HFS choosing which time slot in a cycle to transmit a packet does not affect the time slot selection for other packets of the same flow. As HFS gets rid of the coupling relationship in determining  the transmission time slot among packets of FCS, HFS gets more scheduling options. 
This experiment reaffirms our analysis in section \ref {sec:HFS_schedule_analysis}. 
In particular, in this experiment, we see that HFS  schedules 6$\times$ the number of flows as FCS does.

\subsubsection{Gain in Number of Feasible Solutions}
To explore the increase in the solution space of HFS over FCS, we evaluate the number of feasible solutions while considering a single flow in the network. In both the AFDX and Ladder topologies, for generality, we generate 72 and 56 flows with their length of  cycle randomly selected from the set $\{10,15\}$. We count the number of each flow's  feasible solutions to draw the corresponding box plot. The hypercycle contains 30 time slots. For each generated flow, we use ${Networkx.all\_simple\_paths()}$ in Python to count the number of simple paths in the lifespan of a packet of the flow as the number of feasible solutions under FCS. This is true since the lifespans of different packets do not overlap (because the lengths of their cycle and allowed delay are the same). Regarding HFS, for each packet of a flow within a hypercycle, we count the number of simple paths within the packet's lifespan. Next we multiply the number of feasible paths of each packet to obtain the number of feasible solutions of HFS.

\begin{figure}[htbp]
    \centering
    \subfigure[AFDX]{
    \centering
    \includegraphics[scale=0.6]{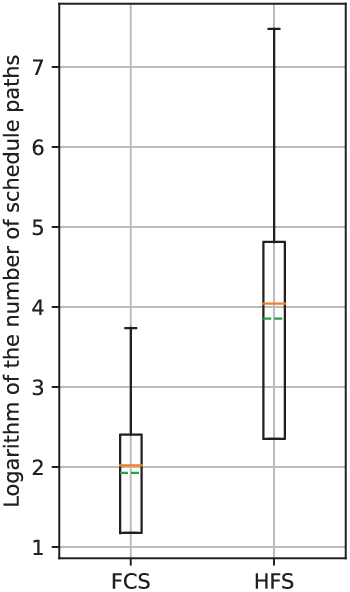}
    }
    \subfigure[Ladder]{ 
    \centering
    \includegraphics[scale=0.6]{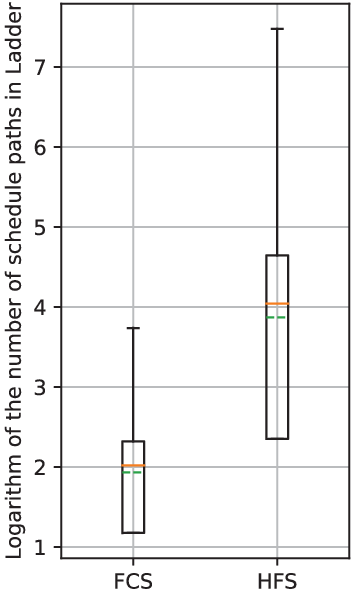}
    }
    \caption{Box plot denoting the number of feasible solutions for FCS and HFS schemes on AFDX and Ladder topologies.}
    \label{fig:boxplot_solution_space}
\end{figure}

Fig. \ref{fig:boxplot_solution_space}(a) and (b) plot the 
size of the solution space for the two scheduling schemes, FCS and HFS using a box plot, in AFDX and Ladder topology, respectively. The 
yellow line denotes the median of the values, the green dotted line denotes the mean value, and the whiskers extend to at most 1.5 times the interquartile range from the upper and lower quartiles. In the shown Figure, no lower whiskers exist for both box plots meaning the  25\% of flows have the same number of feasible solutions. 
Notably, in both networks, HFS can increase the number of feasible solutions of FCS by 100 times on average, while increasing the maximum number of schedule paths by thousands of times, showing the much stronger schedulability of HFS as compared to FCS. Such gain comes from the freedom to choose paths within each cycle of a flow instead of 
fixing the same path in each cycle.   

\subsubsection{General Performance Gain}
\begin{figure}[htbp]
    \centering
    \subfigure[AFDX]{
    \centering    \includegraphics[scale=0.23]{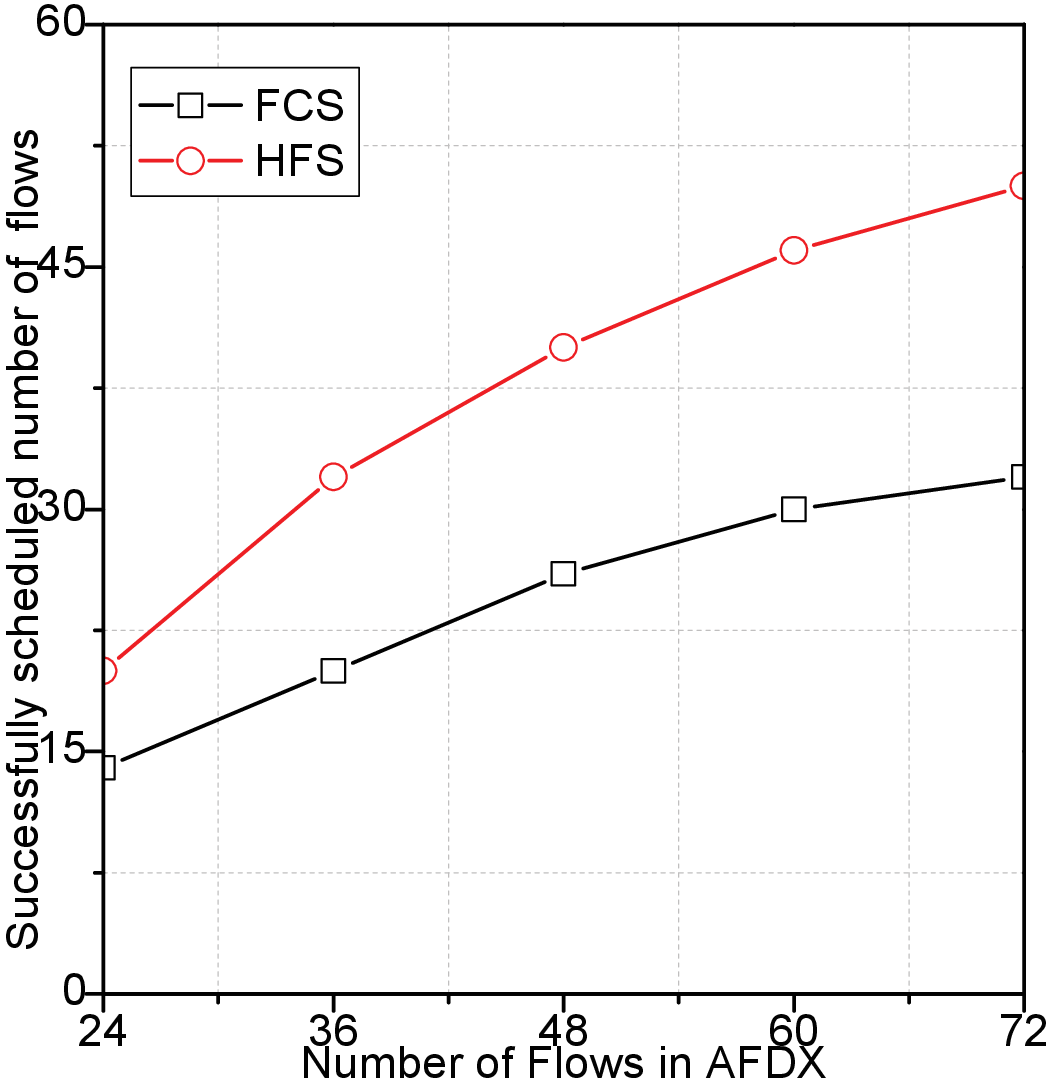}
    }
    \subfigure[LADDER]{ 
    \centering    \includegraphics[scale=0.23]{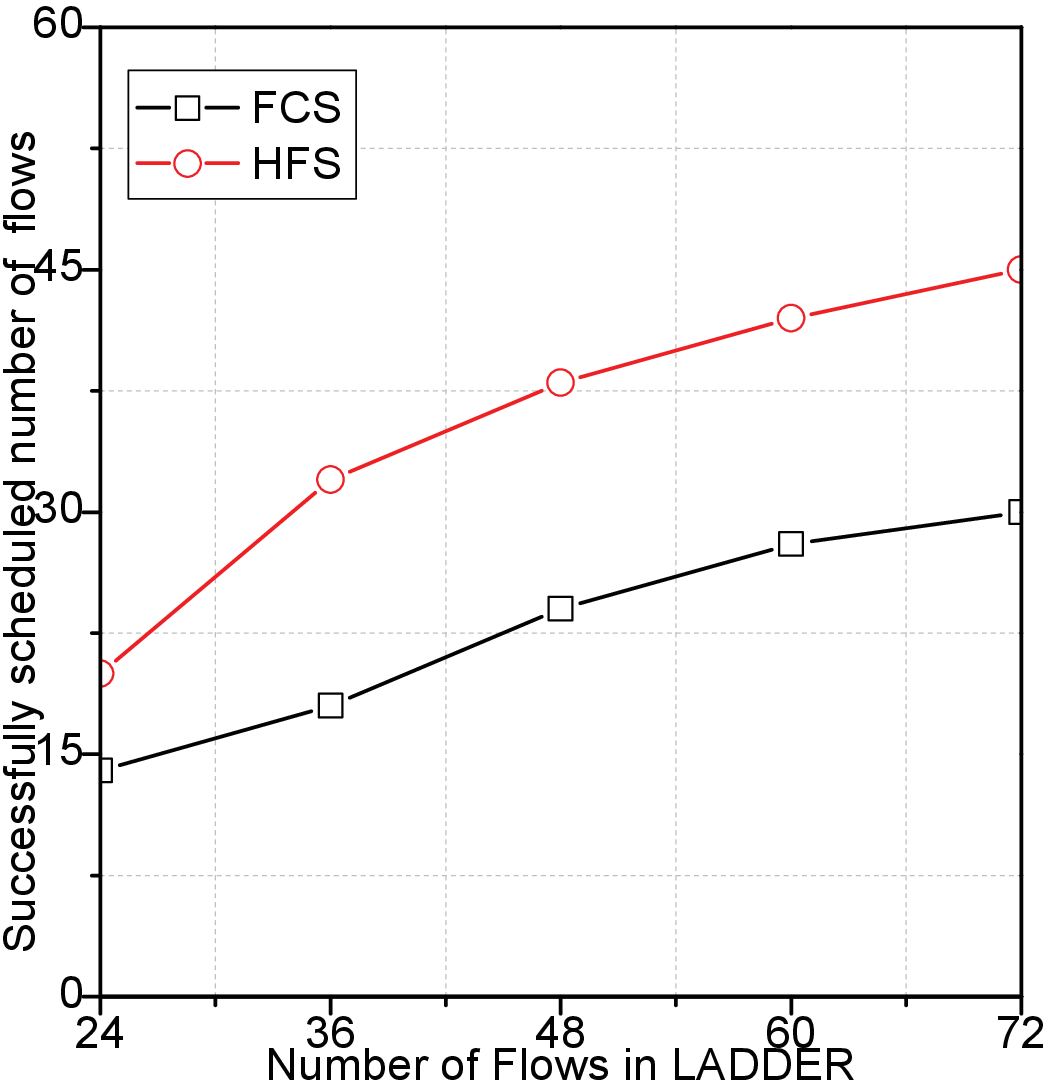}
    }
    \caption{The number of successfully scheduled flows by HFS and FCS schemes on AFDX and Ladder topologies. }    \label{fig:fcs_vs_hfs}
\end{figure}

Fig. \ref{fig:fcs_vs_hfs} shows the performance gain of HFS over FCS under AFDX and LADDER topology, as the number of given flows increases from 24 to 72. 
Each flow was randomly assigned a cycle length of either 5 or 6. In particular, we observe that HFS can schedule $1.5-1.6\times$ the number of flows scheduled by FCS under AFDX topology and can schedule $1.5-1.78\times$ the number of flows scheduled by HFS under LADDER topology. 
Notice that, since we are trying to schedule 2 flows with co-prime cycle lengths, according to \textit{Corollary \ref{corol:co-prime}}, it would imply a theoretical gain of 2 times. The current experiment, however, does not achieve this theoretical gain (i.e., 2 times gain) due to the presence of possible edge-disjoint paths for flows with co-prime cycles. That is, to some extent, FCS can remove the incompatibility between flows with co-prime cycles by assigning them edge-disjoint paths.


\subsubsection{Effects of Tolerable Delay on Performance Gain}

\begin{figure}[htbp]
    \centering
    \subfigure[AFDX]{
    \centering    \includegraphics[scale=0.23]{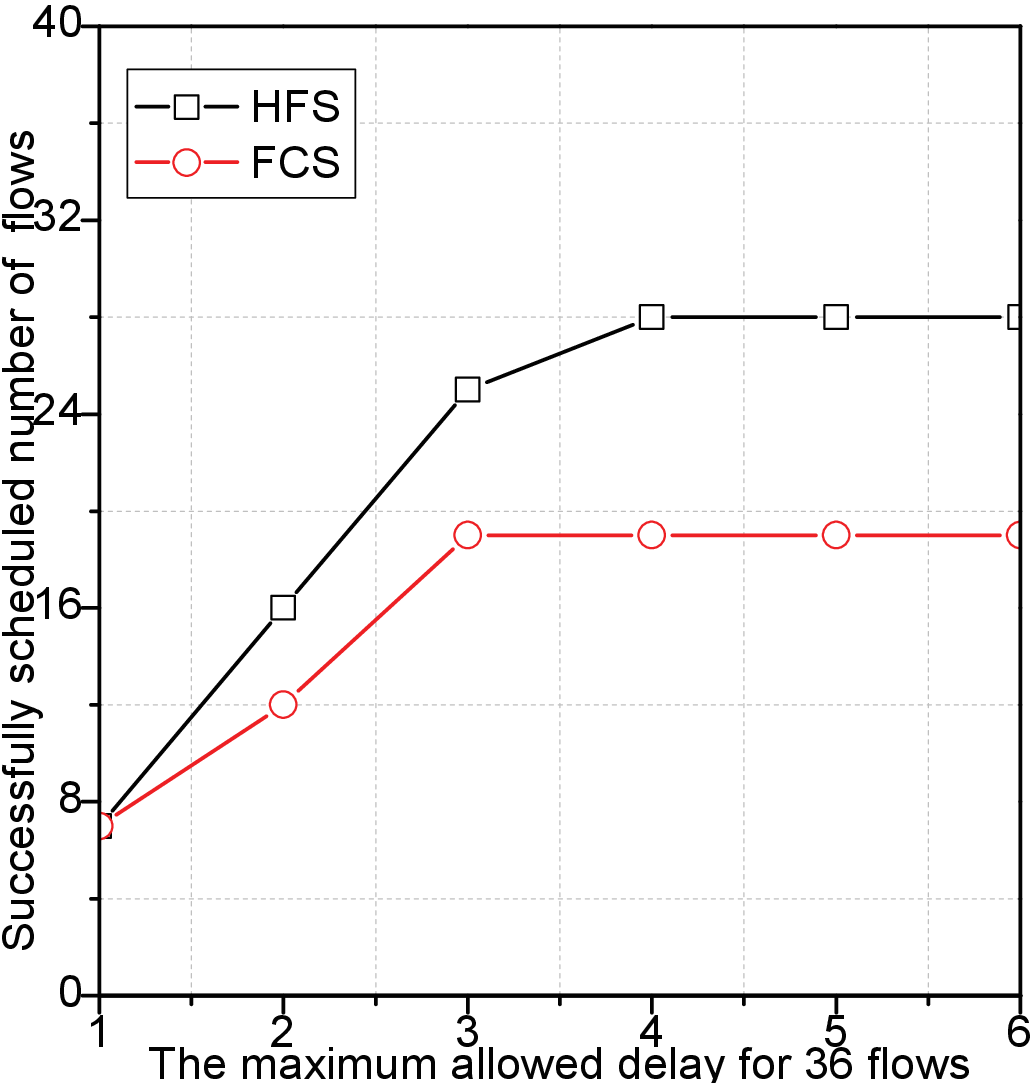}
    }
    \subfigure[LADDER]{ 
    \centering    \includegraphics[scale=0.23]{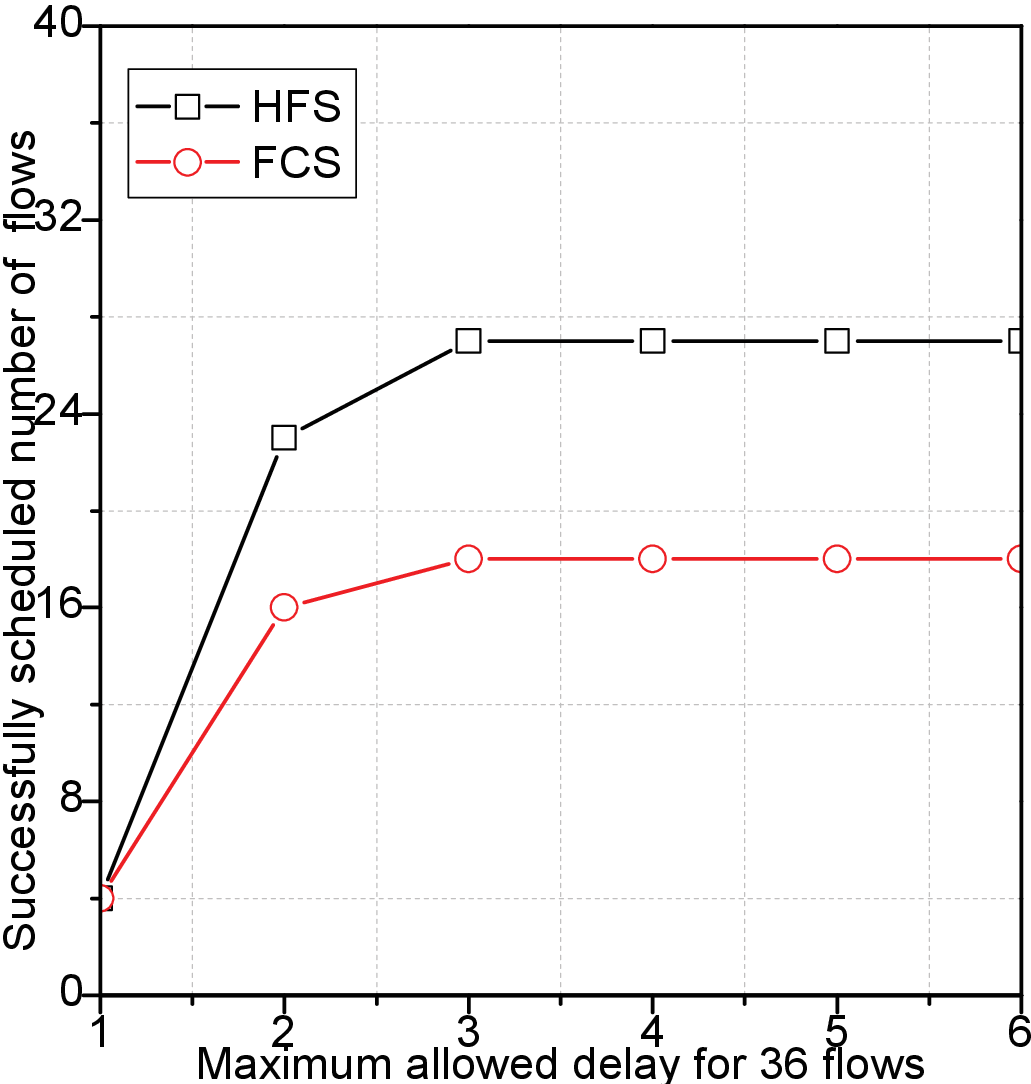}
    }
    \caption{The effects of different maximum allowed delay on the performance in AFDX and LADDER. }    \label{fig:delay_gain}
\end{figure}

We conduct experiments under the AFDX and LADDER topologies to evaluate the impact of different maximum allowed delays on the performance gains of HFS compared to FCS. The results are presented in Fig. \ref{fig:delay_gain}(a) and Fig. \ref{fig:delay_gain}(b). In both networks, we consider 36 flows, categorized into two types with cycle lengths of 5 and 6 time slots, respectively. The given maximum allowed delay is identical for both flow types.

As the  maximum allowed delay increases from 1 to 6 time slots, both HFS and FCS admit a greater number of flows. This behavior aligns with expectations, as a higher maximum allowed delay provides more flexibility in path selection for each packet. Specifically, in the AFDX topology, as the maximum allowed delay increases from 1 to 4 time slots, the number of flows admitted by HFS increases by a factor of 4$\times$, compared to 
2.71$\times$ for FCS. Similarly, in the LADDER topology,  as the maximum allowed delay increases from 1 to 3 time slots, HFS admits 
6.75$\times$ more flows, while FCS admitted 4.5$\times$ more flows. These results indicate that HFS benefits significantly more than FCS from increased maximum allowed delays, primarily due to its ability to utilize additional communication capacity more effectively. That is, unlike FCS, which requires identifying periodic paths for all packets of a flow in the time-expanded cyclic graph, HFS only needs to identify a viable path for each packet within a hypercycle.  

Furthermore, as depicted in Fig. \ref{fig:delay_gain}(a), in the AFDX topology, the growth of the number of flows admitted by increasing the given maximum allowed delay, stagnates earlier in FCS as compared to HFS. Specifically, as the maximum allowed delay increases from 3 to 4 time slots, FCS cannot take advantage of the additional communication resources of TTEthernet to admit more flows, while HFS successfully utilizes these resources to accommodate more flows.

Overall, both HFS and FCS benefit from an increase in the maximum allowed delay. We see that the maximum performance gain of HFS over FCS is achieved when the maximum allowed delay is less than the cycle length in both the AFDX and LADDER topologies. Notably, the point at which HFS achieves its optimal performance gain over FCS does not necessarily exceed the cycle length; rather, it is influenced by the underlying network structure.


\subsection{Comparison between HFS-LLF and Optimal Scheme}
In this subsection, we evaluate the runtime as well as the number of successfully scheduled flows of HFS-LLF as compared to HFS to show the effectiveness of HFS-LLF. We conduct these experiments on AFDX and LADDER topologies since it becomes intractable for the solver to solve HFS on large-scale networks.

\subsubsection{Runtime}
\begin{figure}
\centering
\includegraphics[scale=0.25]{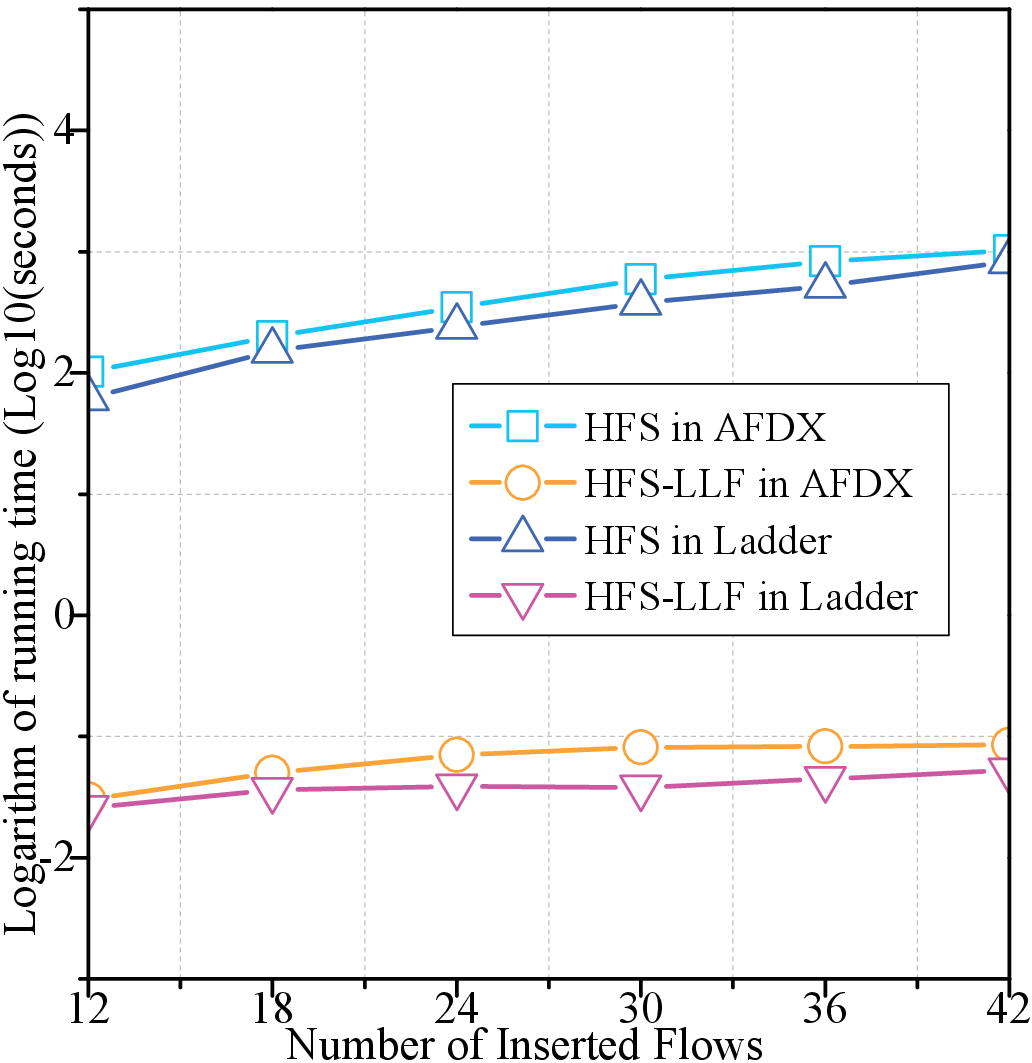}
\caption{The running time of HFS and HFS-LLF given different flows on the AFDX and Ladder topologies.}
\label{fig:runtime}
\end{figure}

Fig. \ref{fig:runtime} plots the running time of HFS and HFS-LLF as the number of generated flows' increases. For each flow, its cycle length is chosen randomly from the values $\{2,3,5\}$ with the hypercycle length being 30. In both topologies, the runtime of HFS increases by around 10 times as the number of arrived flows increases from 12 to 42, while the runtime of HFS-LLF only increases by around 2.2 times. Particularly, the running time of HFS can reach $10^4=10000$ times that of HFS-LLF, highlighting the efficiency of the graph-based HFS-LLF algorithm. This gain comes from HFS-LLF exploring the special structure of this problem using a shortest-path algorithm.   

\subsubsection{Optimality Gap}
Table \ref{tab:scheduled_flows} shows the number of successfully scheduled flows of HFS and HFS-LLF as the number of arriving flows increases from 18 to 54 (in steps of 6). As before, each flow's cycle length is chosen randomly from the values $\{2,3,5\}$ with the hypercycle length being 30. The first row of table \ref{tab:scheduled_flows} shows the given number of flows while the second and third rows show the successfully scheduled number of flows and the last row shows the gap between our HFS-LLF scheme and the optimal scheme (i.e., HFS). Note that, no polynomial scheme can be optimal due to the NP-Hardness of the problem. Compared to our proposed graph-based scheme HFS-LLF, the biggest observed gap between our scheme and the optimal value is less than $9.7\%$.

\begin{table}[htb]
\caption{Number of Successfully scheduled flows in AFDX}
\label{tab:scheduled_flows}
\centering
\begin{tabular}{@{}c|ccccccc@{}}
\toprule
\rowcolor[HTML]{96FFFB} 
Number & \multicolumn{1}{c|}{\cellcolor[HTML]{96FFFB}18} & \multicolumn{1}{c|}{\cellcolor[HTML]{96FFFB}24} & \multicolumn{1}{c|}{\cellcolor[HTML]{96FFFB}30} & \multicolumn{1}{c|}{\cellcolor[HTML]{96FFFB}36} & \multicolumn{1}{c|}{\cellcolor[HTML]{96FFFB}42} & \multicolumn{1}{c|}{\cellcolor[HTML]{96FFFB}48} & 54 \\ \midrule
HFS & 13 & 17 & 19 & 21 & 24 & 31 & 34 \\
HFS-LLF & 12 & 16 & 18 & 20 & 22 & 28 & 31 \\ \midrule
Gap & 7.7\% & 5.9\% & 5.3\% & 5\% & 8.3\% & \cellcolor[HTML]{FD6864}9.7\% & 8.8\% \\ \bottomrule
\end{tabular}
\end{table}

The experiments here show that our proposed HFS-LLF has a comparable throughput to that of HFS but with a significantly faster running speed. 
We further compare HFS-LLF to other existing schemes with a lager size of problem in the following.  

\subsection{Evaluation of HFS-LLF on AFDX and LADDER}
\begin{figure}[htbp]
    \centering
    \subfigure[AFDX]{
    \centering    \includegraphics[scale=0.225]{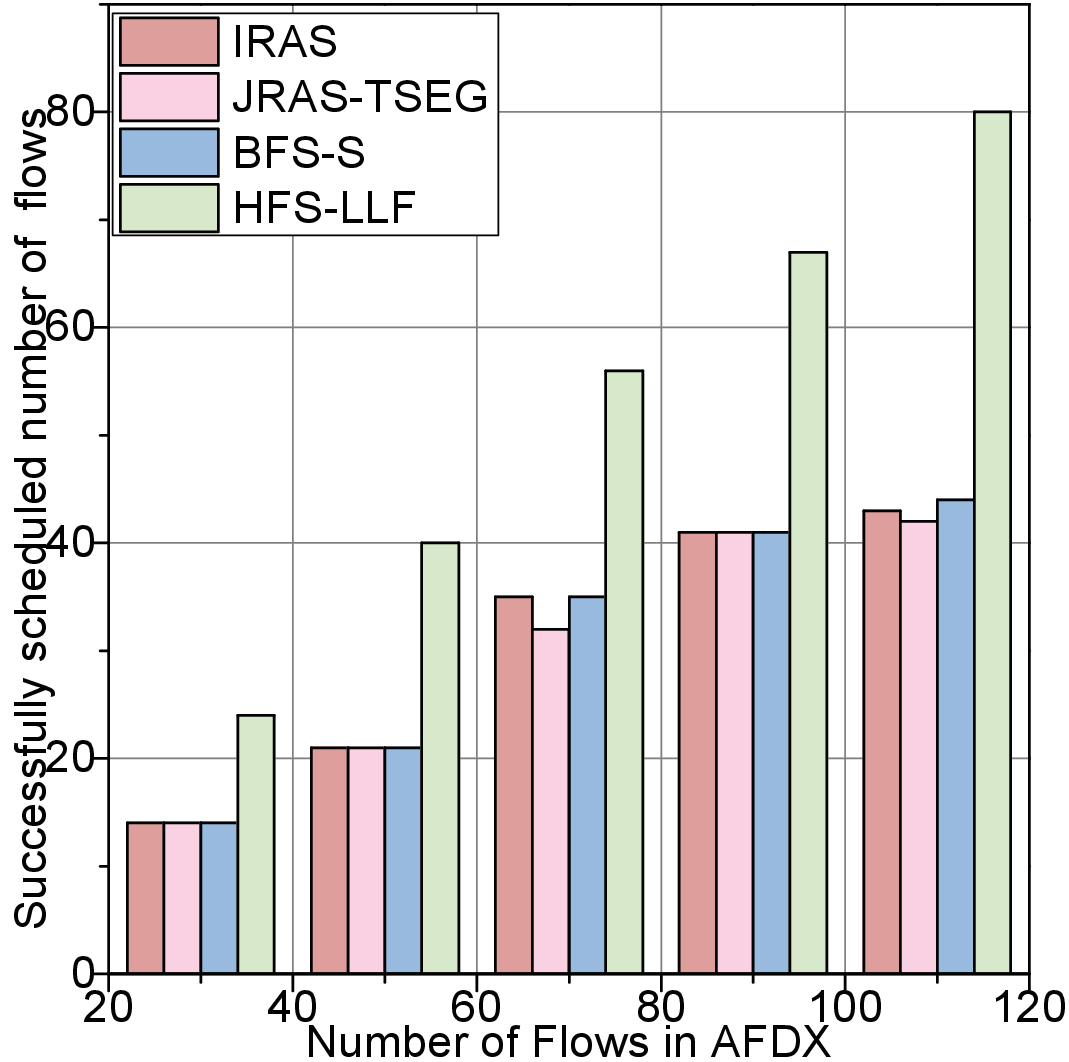}
    }
    \subfigure[LADDER]{ 
    \centering    \includegraphics[scale=0.225]{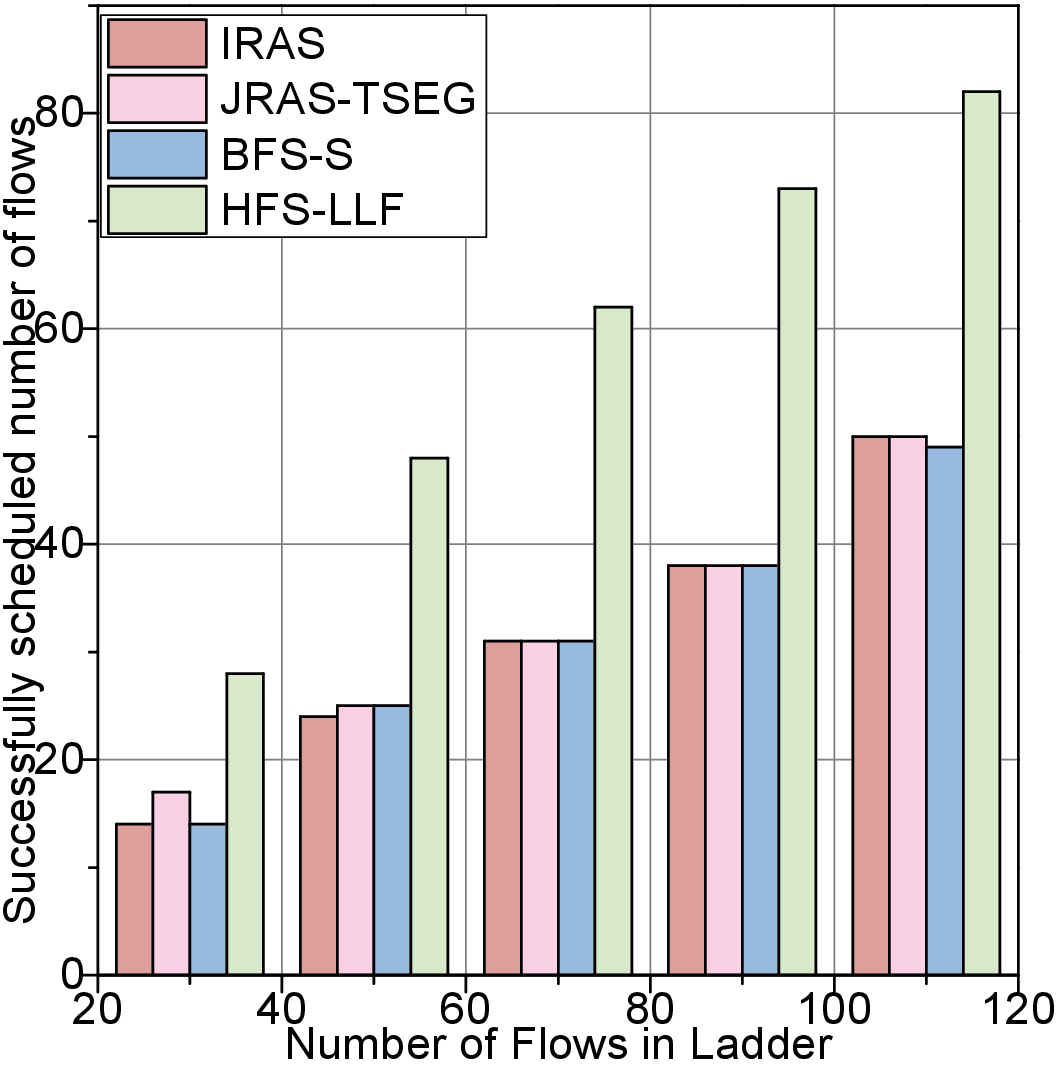}
    }    \caption{The number of successfully scheduled flows for different evaluated schemes on the AFDX and Ladder topologies.}    \label{fig:hfs_llf_afdx_ladder}
\end{figure}

We compare the performance of HFS-based scheme (i.e., HFS-LLF) with the FCS-based schemes~(i.e., IRAS, BFS-S and JRAS-TSEG) on the AFDX and LADDER topologies, respectively, with the number of flows increasing from 30 to 110 with a step of 20. The hypercycle length being 400. Half of the flows in each case have their cycle length as 16 and the other half has its cycle length as 25.
As shown in Fig. \ref{fig:hfs_llf_afdx_ladder}, HFS-LLF outperforms the other 3 schemes by up to $95\%$ and $100\%$  in AFDX topology and LADDER topology, respectively, in terms of the successfully scheduled number of flows. As compared to experiment results in Fig. \ref{fig:fcs_vs_hfs}, a larger performance gain is obtained here. This is due to the cycle length of the flows, i.e., $\{16, 25\}$, are larger than the cycle lengths of $\{5, 6\}$, which requires fewer resources over a single link. Thus, one link can accommodate significantly more flows in HFS scheduling than FCS scheduling, {since HFS can admit flows whose cycle lengths are co-prime in a link while FCS cannot. }

\subsection{Evaluation of HFS-LLF on Random Graphs}
In this subsection, we evaluate the performance of HFS-LLF and 3 other baselines IRAS, JRAS-TSEG and BFS-S in a random graph with heavy load (in terms of the number of input flows), to show the effectiveness of HFS-LLF. In this subsection, the cycle lengths of flows are randomly selected from the set $\{3,4,5\}$ and the hypercycle length is 60. The number of flows with cycle lengths 3, 4, and 5, are equal. The arrival time of each flow is randomly selected from [1,60]. 

\begin{figure}[htbp]
    \centering
    \subfigure[]{
    \centering
    \includegraphics[scale=0.225]{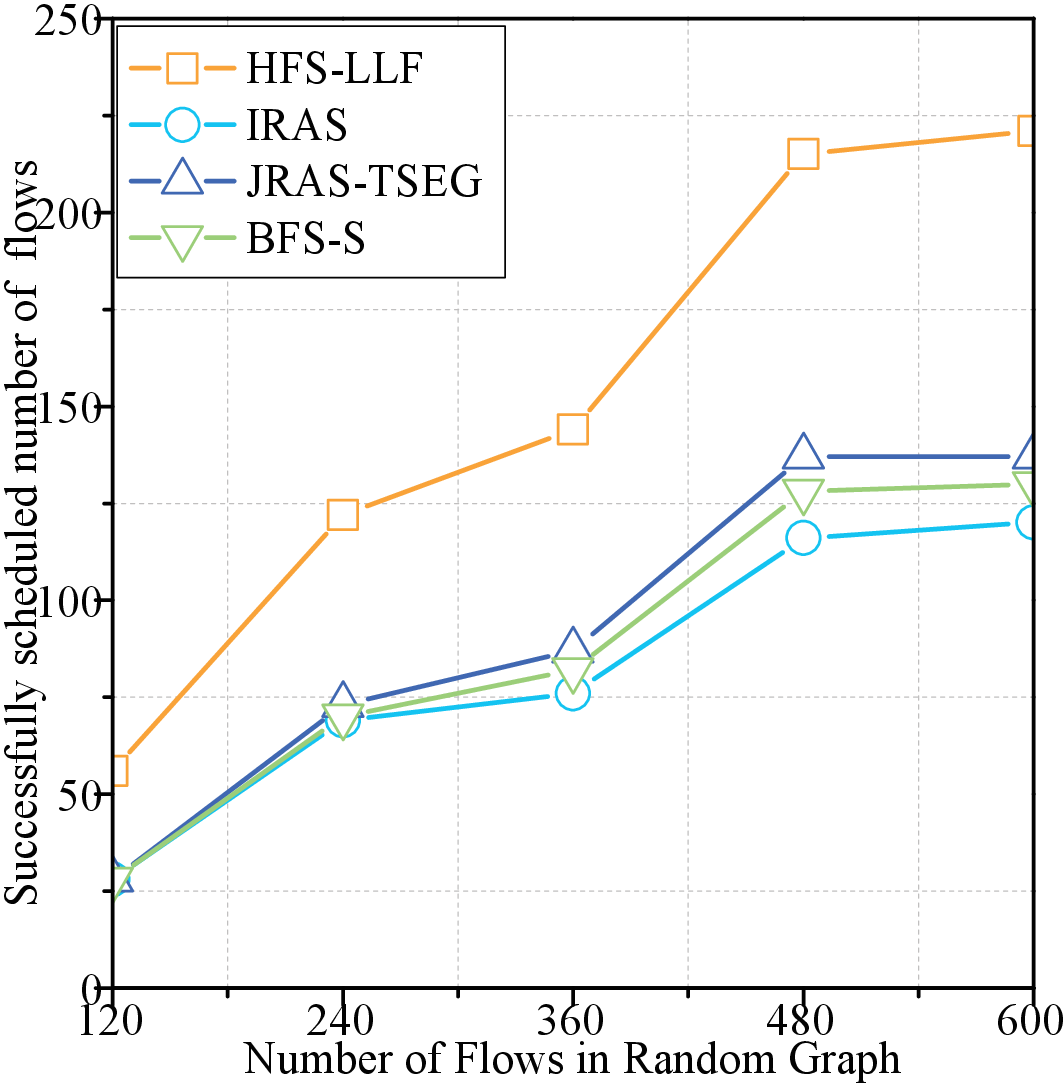}
    }
    \subfigure[]{ 
    \centering
    \includegraphics[scale=0.225]{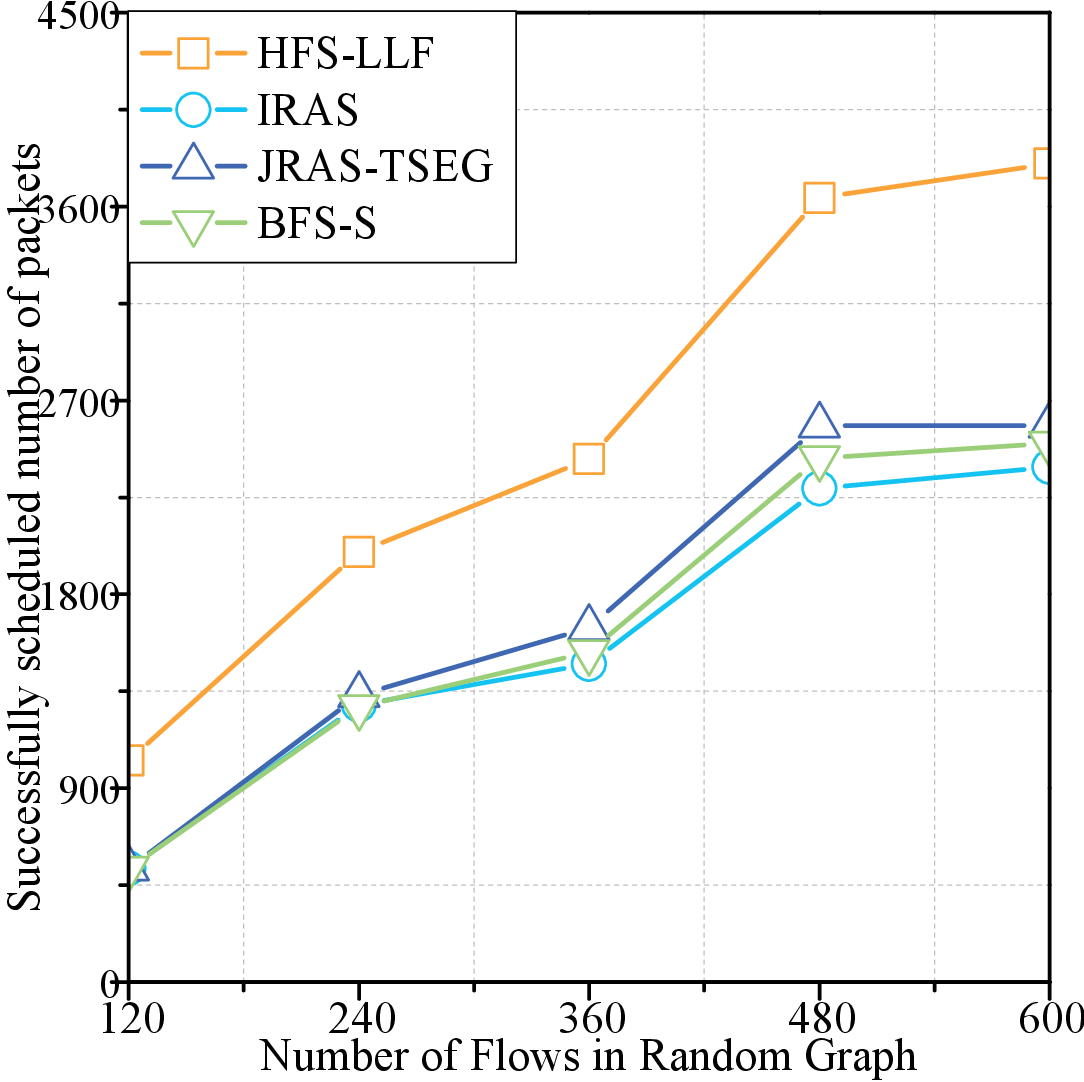}
    }
    \caption{Evaluation of different schemes on a random graph with varying number of input flows. }
    \label{fig:random_graph_load}
\end{figure}

\begin{figure}[htbp]
    \centering
    \subfigure[]{
    \centering
    \includegraphics[scale=0.225]{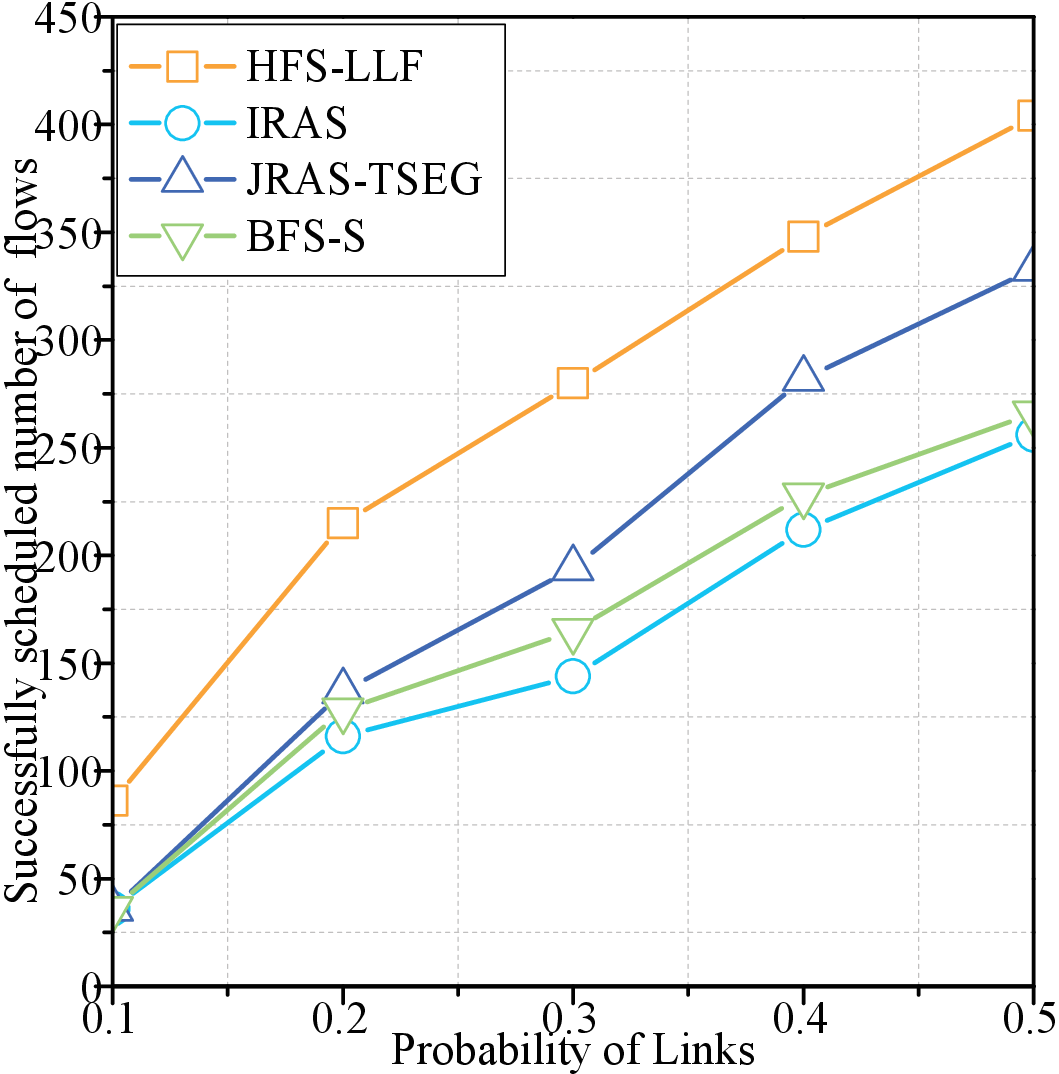}
    }
    \subfigure[]{ 
    \centering
    \includegraphics[scale=0.225]{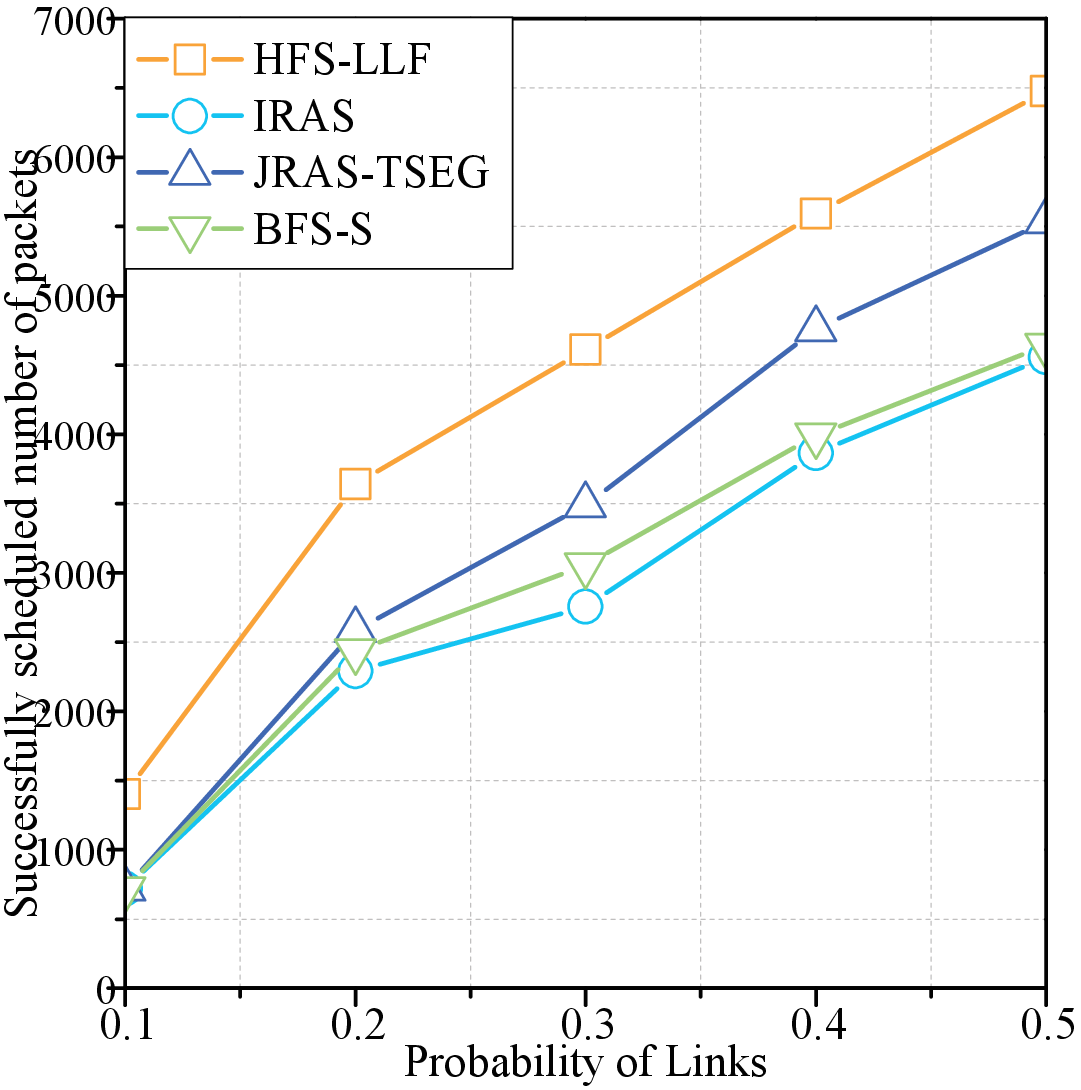}
    }
    \caption{Evaluation of different schemes in a random graph with different link establishment probability.}    \label{fig:random_graph_probability}
\end{figure}

\subsubsection{Effect of different load on performance}
Fig. \ref{fig:random_graph_load}(a) and Fig. \ref{fig:random_graph_load}(b) respectively show the successfully scheduled number of flows and packets of the 4 schemes on a random graph with the probability to generate each link in the random graph as 0.2. As the number of input flows increases, the successfully completed number of flows and packets for each scheme grows rapidly at first and then levels off, which is as expected since the communication capacity almost runs out when the number of flows increases to some extent (i.e., 480). Besides, all the other 3 schemes prevail over IRAS even if IRAS enumerates all the feasible paths to schedule each flow. This is because the other 3 schemes allow the nodes to  temporarily store data while IRAS is a no-wait scheme. As shown in Fig. \ref{fig:ehpg}, a no-wait scheme means removing all the storage edges while the other 3 schemes keep the storage edges. Thus, HFS-LLF, JRAS-TSEG, and BFS-S clearly have more selections of paths than IRAS with more edges in TECG. Moreover, JRAS-TESG outperforms BFS-S since JRAS-TESG considers the suitability of time slots to each link when choosing paths while BFS-S greedily assigns the time slots by simply identifying the smallest-hop path in TECG. Notably, HFS-LLF outperforms the best-performing baseline \textemdash JRAS-TSEG by at least 60\% and at most 100\%
in terms of the number of successfully scheduled flows. Additionally, HFS-LLF schedules  $1.5\times$ to $1.9\times$ the number of packets scheduled by JRAS-TSEG. This is because HFS-LLF can allow any link to accommodate flows with co-prime cycle lengths while such flows with co-prime cycles will be incompatible over a single link under JRAS-TSEG.

\subsubsection{Effect of network size on the performance}
Fixing 480 randomly generated flows as input, we increase the link establishment probability on a random graph from 0.1 to 0.5 to evaluate the successfully scheduled number of flows and packets of the 4 schemes as shown in Fig. \ref{fig:random_graph_probability}(a) and Fig. \ref{fig:random_graph_probability}(b), respectively. 

As the edge density of the random graph increases, the network has a larger communication capacity, which explains why the number of scheduled flows and packets of all the schemes rapidly increases. Notably, as the number of edges increases in the random graph, the performance gap between HFS-LLF and JRAS-TSEG shrinks. When link creation probability $p=0.1$, HFS-LLF schedules $2.4\times$ number of flows scheduled by JRAS-TSEG. This gain decreases to $1.21\times$ when $p$ increases to 0.5. This is as expected since more edges generated can bring more link-disjoint paths into the network. JRAS-TSEG can assign incompatible flows into different edge-disjoint paths to eliminate the conflict. 
We observe that, the number of successfully scheduled packets show a similar trend to the number of successfully scheduled flows since the number of packets of different flows within a hypercycle does not vary much.
Fig. \ref{fig:random_graph_probability} highlights the substantial advantage of HFS-LLF in scheduling heavy load as compared to the fixed cyclic scheduling schemes, since it can remove the incompatibility of flows on a link to make full use of the link resources.  

\section{Discussion}\label{sec:discuss}
\subsection{Dealing with varying packet size
}\label{sec:time_slot}

In scenarios where packet sizes vary significantly,  we can choose a suitable time slot length, for which most of the smaller packets can fit in (without too much waste). For each large packet,  which cannot fit in one time slot, we can fragment the packet   into a smallest number of  packets with equal size  such that each of the fragments can fit in one time slot and the waste can be minimized.   Such a time slot also can only accommodate one packet    With such a time slot,  the bandwidth waste in each time slot can be the round-off to the small packet size and the extra overhead for header of the fragmented packet. Through carefully choosing the length of the time slot that can  balance between the wastage inside a slot for smaller packets and the fragmentation overhead for larger packets, the resource waste can be small. 

 The introduction of the fragmented packets into TTEthernet does not change the problem. That is, 
    the fragments periodically arrive and arrive at the same time with the original packet. To ensure no jitter, the destination node should also receive each fragments of the original packet periodically to merge them back into the original packet periodically. Moreover,  each fragments should also experience a delay  no more than the maximum allowed delay of the original packet such that they can be combined no later than the maximum allowed delay.  The fragments of the original packet are not required to be transmitted in consecutive time slots.   As a result,  HFS/HFS-LLF can consider each fragment of the original packet as a new flow. HFS/HFS-LLF need to successfully schedule the whole group of new flows, corresponding to the fragments of the original flow, to successfully schedule the original flow. 

Our existing simulation results already show that HFS can still significantly outperform FCS under such a time slot setting, since we can group the successfully scheduled flows with the same arrival time and cycle length into a large flow. Obviously, HFS can finish much more large flows than FCS.

\vspace{-3mm}
\subsection{Implementing HFS on TTEthernet}\label{sub:implementation_hfs}

\begin{figure*}
\includegraphics[scale=0.55]{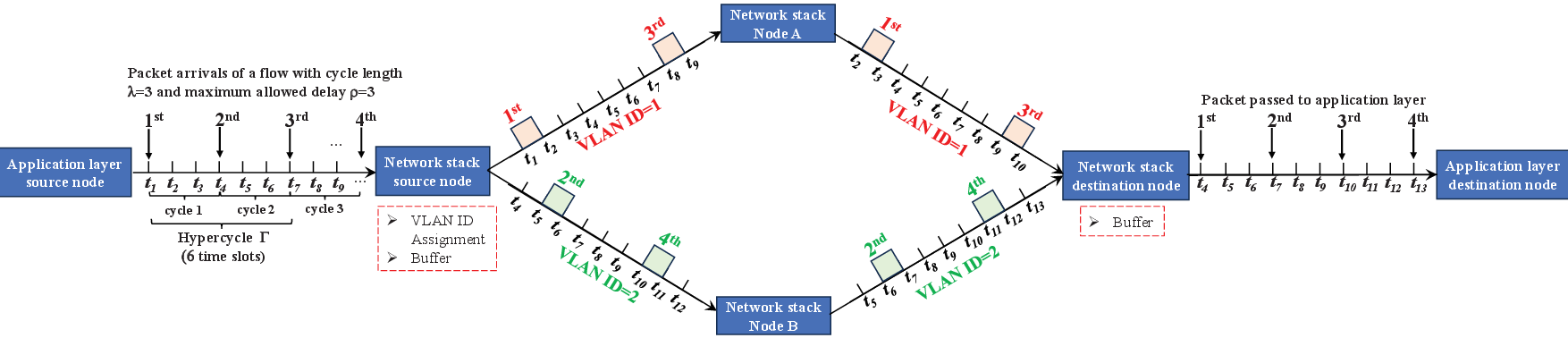}
\caption{An illustration of implementing HFS over TTEthernet using multiple VLANs and destination node buffering}
\label{fig:example_implementation}
\end{figure*}

HFS allows directing the packets within a flow to traverse different paths. To achieve this, 
our HFS scheme can be implemented by mapping different packets of the same flow to multiple different VLANs at the source node and combine them back to the same flow at the destination, We use an example as shown in Fig.  \ref{fig:example_implementation} to illustrate the process, where the network consists of a source node, an intermediate node A, an intermediate node B, and a destination node. Suppose the hypercycle {length} is 6 time slots. Let the application layer of the source node generate a flow with a 3-time-slot maximum allowed delay and a 3-time-slot cycle. The application layer of the source node periodically passes packets to the network stack of the source node on TTEthernet at $t_1, t_4, t_7, t_{10},...$. Take the first 4 packets of the flow as an example,  HFS/HFS-LLF lets the network stack of the source node set the VLAN ID of the 1st and 3rd packets of the flow to 1 while setting the VLAN ID of the 2nd and 4th packet of the flow to 2, so that they can be switched across different paths.
Moreover, the network stack of destination node will respectively receive the  4 packets at $t_3, t_7, t_{10}, t_{12}$, which are not  periodical. The network stack then buffers the 4 packets, according to the maximum experienced delay, i.e., 3 time slots experienced by the 2nd packet. That is, it buffers 1 time slot for the 1st and 4th packets and passes the 4 packets to the application layer respectively at $t_4, t_7, t_{10}, t_{13}$, which are periodical.

Except the VLAN ID assignment at source node and packets aggregation at destination node, the network stacks in HFS do not need to deal with extra processing compared with the network stacks in FCS. As a result, HFS can be implemented without much overhead and using existing networking technologies that are widely deployed.


\vspace{-3mm}
\subsection{Applying HFS to AoI Schemes}

Existing works\cite{aoi_20,aoi_21,aoi_23} also adopt the cyclic schedulers to decide the transmission time slot  of different users  to optimize the AoI. They split the AoI schedule problem into a sub-problem  of deciding the transmission cycle length of different users and another sub-problem of deciding the exact transmission slot for each user.  HFS can potentially reduce the average AoI by supporting more different transmission cycle lengths for different users than other cyclic schedulers (e.g., FCS).

\vspace{-3mm}
\section{Conclusion and Future Work}\label{sec:conclude}

In this work, we explore the throughput gain of hypercycle-level flexible scheduling over the fixed cyclic scheduling. We formulate HFS as an integer linear programming problem and prove that HFS can obtain unbounded performance gain over FCS in terms of the admitted number of flows. We also prove that for a large number of cases with bounded allowed packet delay HFS is optimal and any other flexible scheduling over a larger period than a hypercycle will not outperform HFS. We show that solving HFS is NP-Hard and therefore to efficiently solve HFS, we propose a load status-based greedy algorithm. Simulation results show that HFS can schedule 6 times the number of flows than FCS. Besides, our proposed algorithm HFS-LLF can solve HFS quite fast and with reasonable efficiency. HFS-LLF also outperforms existing schemes by at least 50\% in terms of the number of scheduled flows. In our future work, we will explore the performance gain of  HFS in time-sensitive networks and other deterministic networks.


\balance
\bibliography{RBF}

\end{document}